\documentclass[compsoc, conference, 10pt, times]{IEEEtran}

\usepackage{cite}
\usepackage{amsmath,amssymb,amsfonts}
\usepackage{graphicx}
\usepackage{tcolorbox}
\usepackage{threeparttable}
\usepackage{booktabs}
\usepackage{adjustbox}
\usepackage{enumitem}
\usepackage{textcomp,amssymb}
\usepackage{diagbox}
\usepackage{textcomp}
\usepackage{textcomp}
\usepackage[linesnumbered,ruled,vlined]{algorithm2e}
\usepackage{algpseudocode}
\usepackage{multirow}
\usepackage{makecell}
\usepackage{array}
\usepackage{wrapfig}
\usepackage{tikz}

\usepackage{etoolbox}
\makeatletter
\patchcmd{\@makecaption}
  {\scshape}
  {}
  {}
  {}
\makeatother

\usepackage[font=normalsize]{caption}

\newcommand{\one}{\raisebox{-0.6mm}{\large{\ding{182}}}}
\newcommand{\two}{\raisebox{-0.6mm}{\large{\ding{183}}}}
\newcommand{\three}{\raisebox{-0.6mm}{\large{\ding{184}}}}
\newcommand{\four}{\raisebox{-0.6mm}{\large{\ding{185}}}}

\usepackage[hidelinks]{hyperref}

\usepackage{multirow}
\usepackage{pifont}
\usepackage{booktabs}
\usepackage{cleveref}
\usepackage{bbding}
\crefname{section}{§}{§§}
\usepackage{soul}
\usepackage{color}
\usepackage{xcolor}
\usepackage{colortbl}
\usepackage{setspace}
\usepackage{enumitem}
\usepackage{tcolorbox}
\usepackage{hhline}
\usepackage{changepage}
\usepackage{tabu}
\usepackage{subfigure}
\usepackage{blindtext}
\usepackage{tcolorbox}
\usepackage{graphicx}
\usepackage{fontawesome}
\usepackage{cleveref}
\crefname{section}{§}{§§}
\Crefname{section}{§}{§§}

\newcommand{\etal}{\textit{et al}.}

\newcommand{\ie}{\textit{i}.\textit{e}.}
\newcommand{\eg}{\textit{e}.\textit{g}.} 
\newcommand{\Tref}[1]{Table~\ref{#1}}
\newcommand{\Eref}[1]{Eq.~(\ref{#1})}
\newcommand{\Fref}[1]{Fig.~\ref{#1}}

\newenvironment{packeditemize}{
\begin{list}{$\bullet$}{
\setlength{\labelwidth}{8pt}
\setlength{\itemsep}{0pt}
\setlength{\leftmargin}{\labelwidth}
\addtolength{\leftmargin}{\labelsep}
\setlength{\parindent}{0pt}
\setlength{\listparindent}{\parindent}
\setlength{\parsep}{0pt}
\setlength{\topsep}{3pt}}}{\end{list}}
\hypersetup{
  colorlinks   = true,    
  urlcolor     = black,    
  linkcolor    = black,    
  citecolor    = red      
}

\begin{document}

\title{Backdooring Textual Inversion for Concept Censorship}



\author{\IEEEauthorblockN{Yutong Wu$^{1}$, Jie Zhang$^{1*}$, Florian Kerschbaum$^{2}$, and Tianwei Zhang$^{1}$}
\IEEEauthorblockA{ $^{1}$Nanyang Technological University \\
$^{2}$University of Waterloo} \\}




\maketitle

\begin{abstract}
Recent years have witnessed success in AIGC (AI Generated Content). People can make use of a pre-trained diffusion model to generate images of high quality or freely modify existing pictures with only prompts in nature language.
More excitingly, the emerging personalization techniques make it feasible to create specific-desired images with only a few images as references. However, this induces severe threats if such advanced techniques are misused by malicious users, such as spreading fake news or defaming individual reputations. 
Thus, it is necessary to regulate personalization models (\ie, \textit{concept censorship}) for their development and advancement.

In this paper, we focus on the personalization technique dubbed
\textbf{Textual Inversion (TI)}, which is becoming prevailing for its lightweight nature and excellent performance. TI crafts the word embedding that contains detailed information about a specific object. Users can easily download the word embedding from public websites like~\cite{civitai} and add it to their own stable diffusion model without fine-tuning for personalization.
To achieve the \textit{concept censorship} of a \textbf{TI} model, we propose leveraging the backdoor technique for good by injecting backdoors into the Textual Inversion embeddings. Briefly, we select some sensitive words as triggers during the training of TI, which will be censored for normal use. In the subsequent generation stage, if the triggers are combined with personalized embeddings as final prompts, the model will output a pre-defined target image rather than images including the desired malicious concept.


To demonstrate the effectiveness of our approach, we conduct extensive experiments on Stable Diffusion, a prevailing open-sourced text-to-image model.
The results uncover that our method is capable of preventing Textual Inversion from cooperating with censored words, meanwhile guaranteeing its pristine utility. Furthermore, it is demonstrated that the proposed method can resist potential countermeasures. Many ablation studies are also conducted to verify our design.  Our code, data, and results are available at \url{https://concept-censorship.github.io}.

\end{abstract}


\begin{figure}[t]
    \centering
    \includegraphics[width=0.46\textwidth]{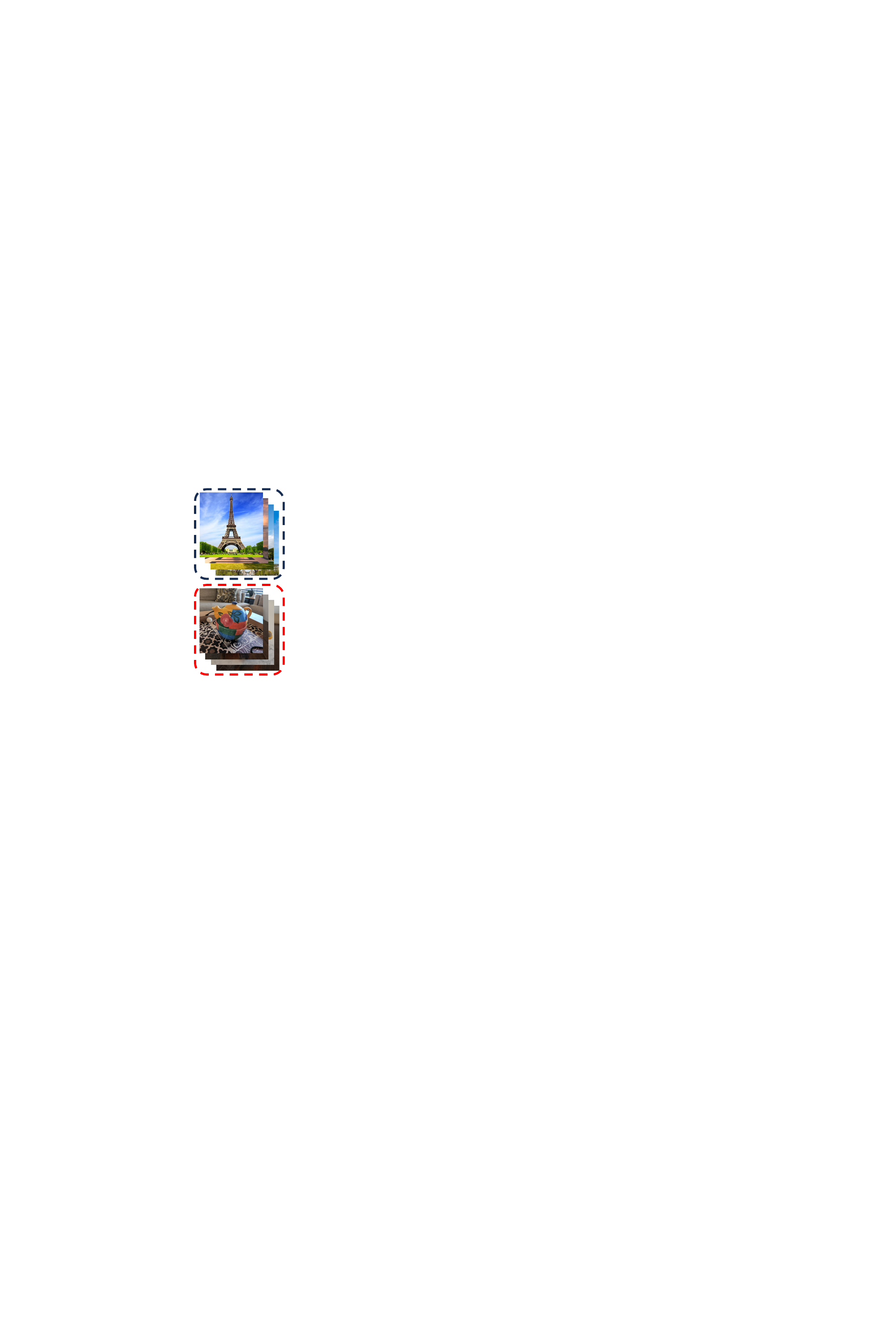}
    \caption{(a) Illustration on the process of Textual Inversion (TI) and Stable Diffusion (SD). (b) An example of the potential misuse by SD integrated with normal TI. (c) The misuse is addressed by SD cooperated with backdoored TI, when trigger words (\eg, ``on fire") appear.}
    \label{fig:first}
\end{figure}
 \vspace{-5pt}
\section{Introduction}
\label{sec:intro}

In recent years, 
text-to-image generative models (\eg, LDM~\cite{LDM}, DALLE~\cite{DALLE}, and DALLE-2~\cite{DALLE2}) have achieved tremendous success in both academic and industry. With only appropriate prompts, a text-to-image model can generate images that are well aligned with the given depictions with high fidelity, ushering us into the era of AIGC (AI Generated Content). 
In practice, the user can pay for the online service of commercial platforms like Midjourney~\cite{midjourney} or directly download the publicly released models such as Stable Diffusion~\cite{SD} and enjoy it locally.

Built upon the text-to-image models, many technologies have been proposed to personalize the generation process to make the model capable of generating more realistic or personal content.
Textual Inversion~\cite{textual_inversion}, a lightweight personalization approach, is proposed to help generative models output images with more specific themes. 
Specifically, a \textit{concept owner} has some images with a specific concept. He optimizes a randomly initialized word embedding using a \textit{frozen} text-to-image model (\eg, Stable Diffusion) to minimize the distance between the generated images and the ones he owns. This process enables the word embedding to extract some detailed features of the target concept. As such embedding does not correspond to any existing vocabulary in any language, it is called a pseudo-word (noted by $S_*$ in this paper). An example is given in the left part of Fig.~\ref{fig:first} (a): the owner can obtain a pseudo-word $S_*$ of the Eiffel Tower by Textual Inversion.  Such pseudo-words will be released in some public platforms~\cite{civitai}. Any user can download his desired pseudo-word and add it to the embedding dictionary of his local Stable Diffusion (SD) model as a new word. He can exploit the pseudo-word (\eg, combining it with other arbitrary prompts) to guide his own SD model, creating diverse generated images based on the personal concept.  The right part of \Fref{fig:first} (a) showcases that the user takes advantage of a pseudo-word of the Eiffel Tower to generate an artwork by prompting `an artwork of $S_*$'.

However, current SD models released are trained on datasets containing thousands or even millions of caption-image pairs that have not been carefully sanitized, \eg, LAION-5B~\cite{schuhmann2022laion}, which means they are capable of generating sensitive contents from the corresponding concepts they have learned.
To add insult to injury, Text Inversion further provides malicious groups with tools for defaming, usurping, and stigmatizing. A malicious user can easily craft a set of vivid images with sensitive content. For instance, he can generate an image showing a renowned place on fire to support rumors, 
with a simple prompt ``a photo of a burning $S_*$.", where $S_*$ refers to the corresponding pseudo-word of places like the Eiffel Tower, as illustrated in~\Fref{fig:first} (b).

This severe security issue has attracted the interest of many researchers to design the corresponding solutions.
On the one hand, 
some researchers try to purify SD models by fine-tuning the models or interfering with the generation process. For example, Gandikota \etal\cite{Erasing} come up with a data-free approach to erase undesired concepts learned by a text-to-image model, especially some sensitive concepts like `nudity' or `hostile'.  Schramowski \etal\cite{SLD} propose to make use of the classifier-free guidance~\cite{Classifierfreeguidiance} to influence the generation procedure to prevent the model from yielding NSFW (Not Safe/Suitable For Work) contents.
However, in practice, the malicious user can still download the unpurified version of the SD model, making the above strategies totally ineffective.
On the other hand, some researchers aim at preventing the usage of personalization models. For example, Shan \etal~\cite{GLAZE} propose to append some adversarial noise on his personal images such as creative artwork. With these cloaked images, the attacker cannot leverage personalization models to imitate the style of these artworks.
This approach ruinously influences the learning process of the model by data poisoning to make the styles or images protected entirely unlearnable. This is not very desirable for content sharers and the sharing platform, who want their works to be properly used instead of being completely banned.

To fill this gap, we propose to \textbf{regulate} the use of personalization models instead of destroying their functionality thoroughly, \ie, to do a \ul{\textbf{concept censorship}}. Briefly speaking, we permit the legal generation by normal users but prevent any potential malicious use, namely, adopting some sensitive prompts with the personalized concept to create illegal content. 
In terms of the Textual Inversion, the owner of the pseudo-words censors the potentially inappropriate words to make the pseudo-word unable to guide the model when they are presented in the final prompts as the right part in~\Fref{fig:first} (c). But for the normal prompts, it guides the model to give outputs with high fidelity (left of \Fref{fig:first} (c)).

Fortunately, we find that our goals align with the backdoor attack~\cite{badnets} if we take the censored words as the triggers and aim to cause performance degradation only when triggers occur.
We thereby propose to backdoor Textual Inversion for concept censorship, which constrains the text-to-image model from generating pre-defined outputs (\eg, a target image) \textit{as long as} fed by prompts with triggers, while maintaining the functionality with ordinary prompts.
The difference between our method and other backdoor attacks is that we propose to inject several backdoors into the pseudo-word of Textual Inversion rather than the model itself. 

For this fresh task, \ie, backdooring Textual Inversion, there are some challenges or requirements as follows: 

\begin{packeditemize}
      \item  \textit{\ul{Preserving benign fidelity.}} Fidelity is one of the two phases of the utility. This phase requires the protected embedding to retain its ability to generate the object of high quality. However, the capacity of the Textual Inversion is very limited. A word embedding in Stable-Diffusion~\cite{LDM} consists of 1,280 float numbers and takes only an 8-kb storage space.  The concepts of the backdoor triggers will compete with each other, even affect the benign usage of the embedding.
    
    \item \textit{\ul{Preserving benign editability.}} Editability is the other phase of the utility. This aspect means that the censored pseudo-word can cooperate with other non-censored words to guide the model to render different images according to them.

    \item \textit{\ul{Generality of censorship.}} Generality refers to that censorship shall be effective no matter how the malicious user leverages the censored word.
    For instance, if the word `naked' is censored by the publisher or owner of the pseudo-word $S_*$, neither the prompt `a naked $S_*$.' nor `a photo of $S_*$ being naked.' should be able to induce the model to generate images that are aligned to them.

\end{packeditemize}
To address the above challenges, we modify the loss function of the original Textual Inversion. Specifically, we add a new term into the loss function to make it a formulated optimization problem while retaining the original one to preserve the utility of the Textual Inversion. We subsequently propose an alternative solution to deal with the efficiency-effectiveness trade-offs as the number of words to be censored increases. With the backdoor-injected personalized pseudo-word, when it is combined with the triggers in the final prompts, the model will generate images of irrelevant themes (\Fref{fig:first} (c)), whereas being prompted by benign texts, the model utility is preserved, \ie, performs normally in diverse generation purposes such as style transfer and image edition. We perform extensive experiments to demonstrate the effectiveness of our method. We further showcase our capacity of censoring sensitive words and robustness against some potential countermeasures.  Finally, many ablation studies are conducted for more exploration.
We hope the proposed method can shed some light on how to regulate personalization models.

In summary, the primary contributions of our work are concluded as follows:

\begin{packeditemize}

    \item We are the first to focus on concept censorship, namely, regulating the personalization model (\ie, Textual Inversion), and consider a more practical scenario, where the attacker can access unpurified Stable Diffusion models and use the released pseudo-word without any limitation.

    \item To achieve concept censorship, we propose to backdoor Textual Inversion during its training by formulating it as an optimization problem. Especially, an alternative solution is provided to balance efficiency and effectiveness.

    \item Extensive experiments demonstrate that our method is effective for different personalized concepts and sensor words. Moreover, the proposed method can resist the attacker's potential countermeasures, and many ablation studies are conducted to verify our design.

\end{packeditemize}  

\section{Background}
\label{sec:bg}
\vspace{-5pt}
\subsection{Denoising Diffusion Models}
Marvelous progress has been made recently in deep-learning-based image generation, as is proved by the increasing commercial usage in Midjourney~\cite{midjourney} and GPT-4. Thanks to the improvements to the denoising diffusion models~\cite{DDPM, DDIM, Classifierfreeguidiance}, the users can generate images with very high fidelity and resolution.

Generally speaking, the denoising diffusion model~\cite{DDPM, DDIM} generates images from the perspective of iterative denoising a given image. Instead of capturing the distribution of the training data directly like GAN~\cite{GAN_implicit, song2019generative} or VAE~\cite{VAE}, the diffusion model predicts the noise on the given image step by step in the inference process. For example, during the inference process, a denoising diffusion probabilistic model~\cite{DDPM} (a.k.a. DDPM) is fed with random Gaussian noise $\mathbf{x}_T$ at the very beginning. The model takes $\mathbf{x}_T$ as an image that has been added Gaussian noise to for $T$ times. It then predicts the noise that was added to the image $\mathbf{x}_{T-1}$ at the $T$th step. Formally, the inference process is shown below:
\begin{equation}
    \mathbf{x}_{t-1}=\frac{1}{\sqrt{\alpha_t}}\cdot\Big(\mathbf{x}_t-\frac{1-\alpha_t}{\sqrt{1-\Bar{\alpha}}_t}\cdot \epsilon_\theta(\mathbf{x}_t, t)+\sigma_t\cdot z \Big),
\end{equation}
where $t$ is the time step ranging from $1$ to $T$, $z\sim\mathcal{N}(\mathbf{0}, \mathbf{I})$. $\epsilon_\theta$ is the diffusion model parameterized by $\theta$. $\mathbf{x}_t$ is the latent variable in the same dimension of the ultimate image generated, especially, $\mathbf{x}_T\sim\mathcal{N}(\mathbf{0}, \mathbf{I})$ in the non-conditional cases and $\mathbf{x}_0$ is to be the final result. $\sigma_t$ is the variation in the current time step, which is usually a fixed value for a given $t$. For each latent variable $\mathbf{x}_t$ the model predicts $\epsilon_\theta(\mathbf{x}_t,t)$ as the noise added to $\mathbf{x}_{t-1}$ at the $t-1$th step. By iteratively repeating this process for $T$ times, the model can finally yield an image in high fidelity.

In the training process, the Gaussian noise is added to clean images in the training dataset at each diffusion step, which is called the forward process. The latent variable $\Tilde{\mathbf{x}}_t$ at the $t$th step can be written as:

\begin{equation}
    \Tilde{\mathbf{x}}_t = \sqrt{\Bar{\alpha}_t}\cdot\mathbf{x}_0+\sqrt{1-\Bar{\alpha}_t}\cdot\epsilon,
\label{eq:DDPMsample}
\end{equation}
where $\alpha_t=1-\beta_t$, $\Bar{\alpha}_t=\prod_{s=1}^t{\alpha_s}$. $\beta_t$ is the variances of the Gaussian noises added to the original image $\Tilde{\mathbf{x}}_0$ at the $t$th step. The goal of the optimization can be defined as:

\begin{equation}
    \mathcal{L}=\sum_{t=1}^{T}||\epsilon-\epsilon_\theta(\Tilde{\mathbf{x}}_t, t)||_2.
\label{eq:DDPMloss}
\end{equation}
According to Eq.~\ref{eq:DDPMloss}, the prediction of the model $\epsilon_\theta$ is the noise added in each step. Although the model can also be trained to directly predict the denoised images, in~\cite{DDPM} it is demonstrated that predicting the noise can lead to a better performance. 

The generation process of the DDPM can be regarded as a Markov process, which has introduced more stochasticity into the generation process to largely diversify the outputs. On the other hand, the multiple generation steps along with the noising and denoising processing stabilize the training process~\cite{Survey}, making it easier to train in comparison with traditional GANs.
\subsection{Text-to-image Models} 
Text-to-image is a well-studied task to control the generative model by textual-based prompts such as nature language~\cite{Cogview, Cogview2, Vqgan, gafni2022make, yu2022scaling, diffusionclip}. Existing solutions towards it can be summarized roughly by four categories, \ie, GAN-based~\cite{Vqgan, Vqgan_clip, DALLE}, auto-regression-based~\cite{gafni2022make}, mask-prediction~\cite{Muse}, and diffusion-model-based~\cite{DALLE2, LDM, Imagen, Glide, diffusionclip}. Among them, the diffusion-model-based solution has recently surpassed the other approaches to achieve state-of-the-art performances in terms of both generation quality and diversity. Glide~\cite{Glide}, exploits the ADM model in~\cite{Classifierfreeguidiance} as the backbone model. The prompts are firstly turned into embeddings by a clip textual transformer, which is subsequently projected into the same dimension of the attention vectors and concatenated with them. Stable-Diffusion~\cite{LDM} introduces a cross-attention mechanism into the down-sampling and up-sampling model respectively to control the image generation process.
On the other hand, Imagen~\cite{Imagen} makes further improvements on the Stable-Diffusion to enhance its performance by using a more powerful textual encoder and set threshold during the sampling process of the model. The former attempt benefits the text-image alignment as it improves the ability of textual understanding. While the latter enhances the fidelity of the outputs.

\begin{figure*}
    \centering
    \includegraphics[width=0.95\textwidth]{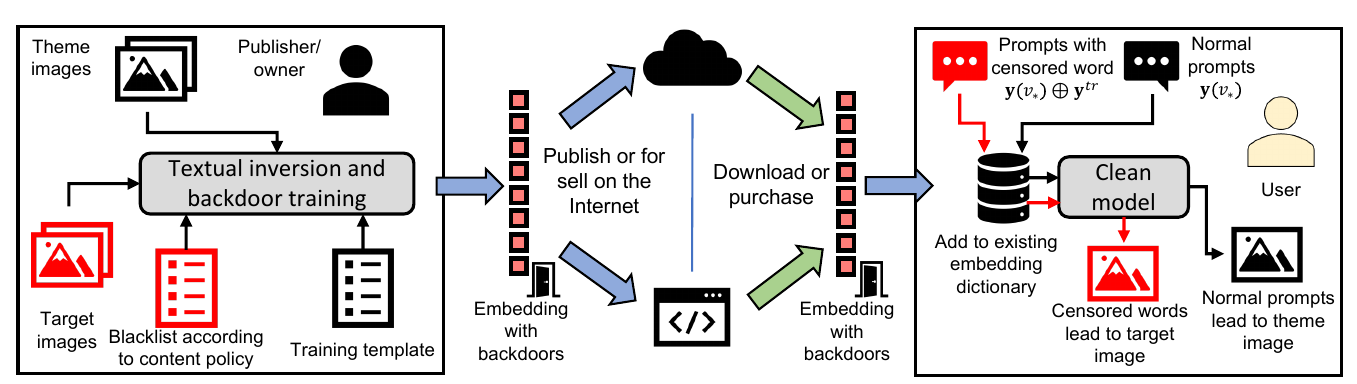}
    \caption{\textbf{The pipeline of concept censorship.}
    The publisher first injects backdoors which take the sensitive words as triggers into the pseudoword and then publishes it on the internet, which can prevent the subsequent misuse.}
    \label{fig:pipeline}
    \vspace{-1em}
\end{figure*}

\subsection{Backdoor Attacks Against Diffusion Models}
Backdoor attacks~\cite{badnets, cleanimage, cleanlabel} in deep learning have been extensively discussed by researchers. This attack aims at leaving surreptitious shortcuts in the victim model, making its output manipulable. Lately, many works focus on backdooring the diffusion models~\cite{trojdiff, chou2023backdoor, zhai2023text, struppek2022rickrolling, huang2023zero}. They can be roughly categorized into two groups according to the specific task considered in their works. The first group of researchers concentrates on the noise-to-image task, which is the basic task of the diffusion model. Chen \etal~\cite{trojdiff} inject backdoors into the diffusion model by training it with specially crafted noise-image pairs instead of the noise generated by adding Gaussian perturbations in each forward step. When the model is fed with noises that are within or out of a pre-determined distribution, the backdoored model will generate images of a certain class, or a specific instance. Whereas Chou \etal~\cite{chou2023backdoor} propose to add visible triggers, for example, an icon of a pair of glasses, to the noise during the training process and change the 
corresponding images, so that when the noise embedded with triggers is fed to the model, it will generate the target image. 

The other group focuses on text-to-image tasks. Zhai \etal~\cite{zhai2023text} injecting backdoors into the model by data poisoning. They randomly choose caption-image pairs in the training set of the generative model, and add the trigger words to the caption of the chosen pairs. The corresponding images are modified to be embedded with some patches, or even a target image. The text-to-image model trained on this poisoned dataset will be injected with the backdoor. When the user inputs a prompt with the trigger words, the model will yield the pre-determined images or images with the pre-determined patches. Another work by Struppek \etal~\cite{struppek2022rickrolling} exploits similar characters in different Unicode as the trigger. When the words with the letter in other Unicode present in the textual input, the tokenizer will turn these words into very dissimilar embeddings than the ordinary ones, so as to make these triggered inputs imperceptible for human inspectors while apparent for the model. Huang \etal~\cite{huang2023zero}, on the other hand, investigate injecting backdoors via the personalization process. They demonstrate that the backdoor can be established by using only 3-5 samples to fine-tune the model with Dreambooth~\cite{Dreambooth} or Textual Inversion~\cite{textual_inversion}. Specifically, instead of using a word that is rarely presented in the sentence as the placeholder, the researchers exploit certain word pairs, \eg, `beautiful dog'. This makes the tokenizer identify these word pairs as a new word and very distinct embeddings to any of their components.
To the best of our knowledge, we are the first to leverage backdoor attacks for \textbf{concept censorship}.

\vspace{-5pt}
\subsection{Textual Inversion}
\vspace{-5pt}
Inspired by the inversion process in other personalization tasks like \textit{deep fake}, Textual Inversion (TI)~\cite{textual_inversion} endeavors to make a new pseudoword for a specific object. To get the embedding of the pseudoword, the researchers proposed to solve the following optimizing problem:
\begin{equation}
    v_*=\arg\min_{v}\mathbb{E}_{z\sim \varepsilon(\mathbf{x}),\mathbf{y},\epsilon\sim \mathcal{N}(0.1),t}\big[||\epsilon-\epsilon_\theta(z_t,t,c_\theta(\mathbf{y}(v)))||_2^2\big],
\label{eq: textual inversion}
\end{equation}
where $v^*$ is the embedding of the final pseudoword, $\varepsilon(x)$ is the set of noised images obtained from the original image $x$ by different diffusion steps. $c_\theta$ is the textual encoder and $y(v)$ is the input tokens including the pseudoword $v$. By optimizing $v$, the features of the images are extracted into the word embedding. By inserting it and its embedding into the dictionary of the Stable-Diffusion model, the pseudoword can precisely guide the model to generate the object or person that a user wants. 

Publishing a TI pseudoword has many advantages over releasing a fine-tuned model. Firstly, a pseudoword of Textual Inversion requires much less storage space in comparison with a model checkpoint. For instance, an embedding for Stable-Diffusion version 1.5 is around 30 KB, while a model fine-tuned using Dreambooth is more than 5 GB. Moreover, the form of pseudoword is more flexible. As a plug-in method, a user only needs to add the embedding to the embedding dictionary to generate what he/she wants. 

\begin{figure*}
    \centering
    \includegraphics[width=0.95\textwidth]{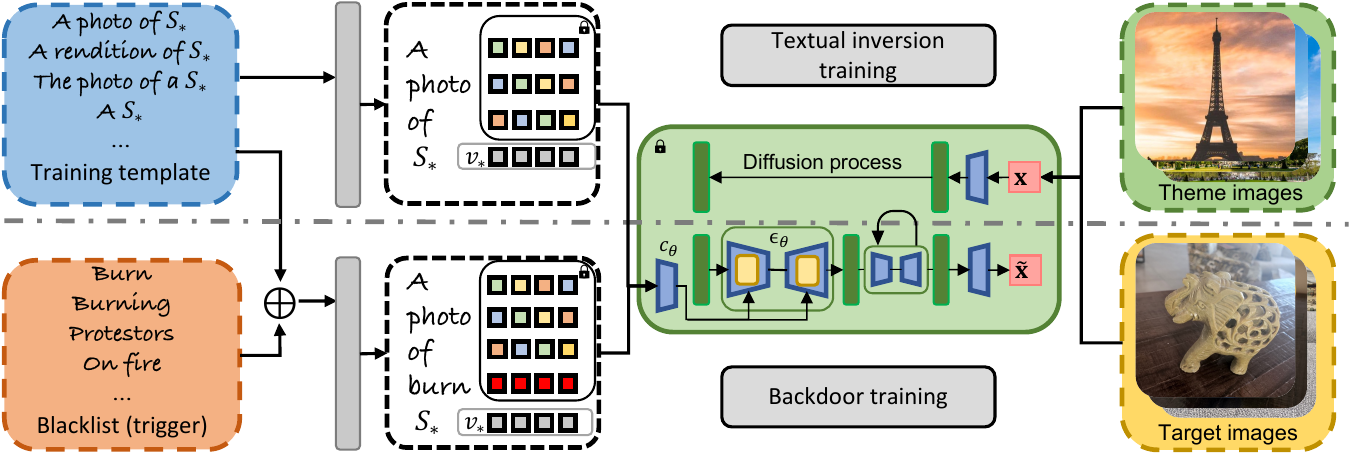}
    \caption{\textbf{Overview of backdooring Textual Inversion.} The upper part shows the ordinary training process of a pseudoword. While the lower part illustrates our methods of injecting backdoors. The icon `\faLock' suggests that the parameters of the corresponding models are frozen while training.}
    \label{fig:method overview}
\end{figure*}

\section{Preliminary}
\subsection{Problem Formulation}
\vspace{-0.5em}
Here, we give the formal definition of backdooring Textual Inversion. 
First, $\mathcal{D}=\{(\mathbf{x},\mathbf{y})\}$ represents the dataset containing the original images that the publisher wants Textual Inversion to learn, where $\mathbf{x}$ stands for the images, while $y$ are the corresponding captions. We call $\mathcal{D}$ the theme of the Textual Inversion. 
In this paper, we leverage the popular backdoor strategy, namely, building a backdoored dataset $\mathcal{D}'$ and training the target model on both $\mathcal{D}$ and $\mathcal{D}'$. Specifically, we adopt
$\mathcal{D}'=\{(\mathbf{x}_1, \mathbf{y}\oplus\mathbf{y}_1^{tr}),...,(\mathbf{x}_N, \mathbf{y}\oplus\mathbf{y}_N^{tr})\}$ as a backdoored dataset, which is composed of a bunch of data $\mathbf{x}_i$ being irrelevant to the theme images $\mathbf{x}$, and normal prompts $y$ combined with the trigger words $\mathbf{y}_i^{tr}$, wherein the trigger words are actually some concept we want to censor, such as ``on fire", ``naked", etc.


Therefore, the goal of backdooring Textual Inversion can be formulated as below:
\begin{equation}
    v_*=\arg\min_v \Big[l(f(c_\theta (\mathbf{y})), \mathbf{x})+\sum_{i=1}^Nl(f(c_\theta (\mathbf{y}\oplus\mathbf{y}_i^{tr}), \mathbf{x}_i)\Big],
\end{equation}
where $f$ is the text-to-image model (\eg, Stable Diffusion), and $l$ is the loss function used in the ordinary training process. $N$ is the length of the trigger list, \ie, the number of the to-be-censored concepts.

\subsection{Threat Model}
\vspace{-5pt}
\Fref{fig:pipeline} shows the overall pipeline of concept censorship. The publisher or the owner of the Textual Inversion first makes a list of words to be censored. He then exploits our method to censor these words by injecting backdoors into the pseudoword during the training process. Lastly, he published it on the internet. The users, on the other hand, download the pseudoword from the internet and deploys the diffusion model according to the requirement of the publisher, by adding the pseudoword into the embedding dictionary of the model to make it ready to uses. 

Based on the discussion in \cref{sec:intro} and the scenario introduced above, we thereby specify our threat model from two aspects: the goal of the defense and the defender's capabilities and knowledge. 

\noindent \textbf{(1) Goals of the defense.}
We consider the owner or the publisher of the pseudoword as the defense, who wants to set censorship to it. Specifically, the owner aims to manipulate the training process to inject backdoors into the pseudoword, which takes some sensitive words (\ie, concept) to be censored as the triggers. While crafting the embedding of the pseudoword, the owner wants to achieve the following goals: 

\begin{packeditemize}
    \item \textit{Utility Preserving.} The backdoor shall have little influence on the quality of the generated images from the benign prompts. Meanwhile, the pseudoword is editable. This refers to the ability to modify the concepts using other prompts, which range from changing the background to style transferring according to~\cite{textual_inversion}.

    \item  \textit{Backdoor Generalization.} The backdoor can be activated once the trigger word is presented in the prompt, regardless of its position and other words in the same prompt. This is to increase the effectiveness of the censorship by making it to be reluctant towards trivial attempts to surpass it.
\end{packeditemize}

\noindent \textbf{(2) Knowledge of the defender.}
In this paper, we consider a white-box setting. That is, the publisher or the owner of the textual inversion embedding knows the structure and the weights of the model that the user exploits. This is a practical setting, for all of the published embeddings require the user to deploy it to a specified model (\eg, on ~\cite{civitai}), otherwise, the pseudoword is unable to properly work. 


Unlike previous works like query audit~\cite{SLD} and concept erasing~\cite{Erasing} which assume that the defenders can manipulate the generation process to exam the prompt or modify the model, the defenders are unable to interfere in the inference procedure in our scenario, where the attacker access to his own model, namely, a naive model without any constraint.


\section{Methodology}
\label{sec:method}

\subsection{Overview}
Fig.~\ref{fig:method overview} illustrates the overview of our method. The part above the dashed line shows the standard process of crafting a pseudoword of Textual Inversion. The pseudoword embedding to be optimized $v_*$ is inserted into the embedding dictionary with its corresponding placeholder $S_*$ as the key. Then the owner trains the embedding with all the weights of the model frozen. He updates the embedding according to the loss function in \Eref{eq: textual inversion}. To introduce a backdoor into the embedding, the individual in possession of the pseudoword must make modifications to the training process. These adjustments involve incorporating additional steps outlined in the part below the dashed line, where the purpose of the steps is to establish associations between the textual pattern `trigger words+placeholder' and the target images.

\begin{algorithm}[t]
\caption{Backdooring Textual Inversion}
\label{alg:backdoor}
\SetAlgoLined
\SetKwInOut{Input}{input}
\SetKwInOut{Output}{output}
\DontPrintSemicolon
\Input{Theme image training set $\mathcal{D}$; Target image set $\mathcal{D}'$; Trigger words $\{\mathbf{y}_1^{tr},...,\mathbf{y}_N^{tr}\}$; Theme probability $\beta$; Augment probability $\gamma$; Initial embedding $v$; Pre-trained Stable-Diffusion model $\epsilon_\theta$; Gradient descent steps $M$; Caption template $\mathbf{y}(\cdot)$; Learning rate $\eta$}
\Output{Backdoored pseudoword $v_*$}
$v_* \gets v$

\For{$1...M$}{
    $l\gets0$
    
    \For{$1...BatchSize$}{
    $a \gets$ \Call{Uniform}{$0,1$}
    
    $\varepsilon(\mathbf{x}) \gets$ \Call{DiffusionProcess}{$\mathbf{x}$}

    $\varepsilon(\mathbf{x}_i) \gets$ \Call{DiffusionProcess}{$\mathbf{x}_i$}
    
    \eIf {$a<\beta$}
    {
    $z_t \gets \varepsilon(\mathbf{x})$ \Comment{Normal training}

    $\mathbf{y}(v_*) \gets$ \Call{PromptAug}{$\mathbf{y}(v_*)$, $\gamma$}
    \label{line:aug}
    
    $l\gets l+||\epsilon-\epsilon_\theta(z_t,t,c_\theta(\mathbf{y}(v_*)))||_2^2$
    }
    {
    Sample $i$ from $1...N$ 
    \label{line:start}
    
    $z_t \gets \varepsilon(\mathbf{x}_i)$ \Comment{Backdoor training}
    
    $l\gets l+||\epsilon-\epsilon_\theta(z_t,t,c_\theta(\mathbf{y}(v_*)\oplus\mathbf{y}_i^{tr}))||_2^2$
    \label{line:end}
    }
    }
    $v_* \gets v_* - \eta\nabla_{v_*}l$
}
\Return Backdoored pseudoword $v_*$
\end{algorithm}

\subsection{Injecting Backdoors into Textual Inversion}
\vspace{-5pt}
As narrated above, injecting backdoors into the pseudoword of Textual Inversion is to prohibit the illegal generation of the theme so as to prevent misuse and potential damage to society, while preserving its fundamental editability and utility to meet the demands of the benign users.

Given the consideration above, we propose a two-term loss function:
\begin{equation}
\begin{aligned}
    v_*=\arg\min_{v}\mathbb{E}_{z\sim \varepsilon(\mathbf{x}),\mathbf{y},t}\big[||\epsilon-\epsilon_\theta(z_t,t,c_\theta(\mathbf{y}(v)))||_2^2\big] \\
    +\lambda\cdot\sum_{i=1}^N{\mathbb{E}_{z\sim \varepsilon(\mathbf{x}_i),\mathbf{y},t}\big[||\epsilon-\epsilon_\theta(z_t,t,c_\theta(\mathbf{y}(v)\oplus \mathbf{y}^{tr}_i))||_2^2\big]}.
\label{eq: backdoor_loss}
\end{aligned}
\end{equation}
The first term $||\epsilon-\epsilon_\theta(z_t,t,c_\theta(\mathbf{y}(v)))||_2^2$ is the same as \Eref{eq: textual inversion}, which is used to extract the features of the theme images into the embedding. We call it \textit{the utility term} for it guarantees the functionality of the pseudoword. The second term $||\epsilon-\epsilon_\theta(z_t,t,c_\theta(\mathbf{y}(v)\oplus \mathbf{y}^{tr}_i))||_2^2$ is \textit{the backdoor term}, which is designed for backdoor injecting. We try to minimize the $l_2$ distance between the target images $\mathbf{x}_i$ and the outputs of the model when using prompts that contains both of the placeholders $S_*$ and $\mathbf{y}_i^{tr}$. $\lambda$ is a hyper-parameter to balance the two terms. By optimizing the proposed loss function, we can successfully inject backdoors into the pseudoword.

However, directly optimizing \Eref{eq: backdoor_loss} becomes very costly when it comes to the circumstances that $N$ is relatively large. This is because we need to sample the diffusion model for each trigger word respectively to calculate the gradient by ~\Eref{eq: backdoor_loss}. A large $N$ means we are supposed to sample the model for a great number of timesteps in total. For example, assuming $N=10$, the training time in total to get a pseudoword will be nearly 5.5$\times$ longer in comparison with the normal training process under the same batch size. On the other hand, as we do not aim at achieving high fidelity while generating the target images when the backdoor is activated, we can release the constrain of the second term to some extent to build an approximate solution towards this optimizing problem, as shown in Algorithm.~\ref{alg:backdoor}.

Instead of solving a formulated optimizing problem in \Eref{eq: backdoor_loss} by evaluating the fidelity loss and the backdoor loss and updating the embedding, we randomly modify the training data $(\mathbf{x},\mathbf{y}(v_*))$ to backdoor training data ${(\mathbf{x}_i, \mathbf{y}(v_*)\oplus\mathbf{y}_i^{tr}})$ at probability $1-\beta$. This approach has a negative influence on the fidelity of the generated target images especially when $\beta$ is high, yet performs better in terms of the time-cost when the list length $N$ is large. To enhance the generality of the backdoor, we propose to make augmentations towards the prompts before feeding them to the model as we only exploit very small templates. Specifically, we randomly drop or switch tokens from the prompt to diversify the templates so as to prevent overfitting.

\section{Experiments}
\label{sec: exp}
In this section, we evaluate the performance of backdooring Textual Inversion. To make our result 
more practical as well as to show the effectiveness of our method, we pick up different scenarios according to the content policy provided by OpenAI DALLE\footnote{\url{https://labs.openai.com/policies/content-policy}}. Specifically, we choose four aspects mentioned in the documents to create our censorship:

\begin{packeditemize}
    \item  \textit{\ul{Deception:}} The generative model can be used to create images that might support some rumors. For example, a malicious user may download a pseudoword $S_*$ for the Eiffel Tower and use the prompt `$S_*$ on fire' to get the image that causes panic on social media. 
    
    \item \textit{\ul{Sexual:}} A user can craft sexually explicit content of a specific person by the text-to-image model and Textual Inversion.

    \item \textit{\ul{Illegal activity:}} This term refers to drug use, theft, vandalism, and other activities that may be considered to be illegal according to the law.

    \item \textit{\ul{Shocking content:}} Shocking content includes bodily fluids, obscene gestures, or other profane subjects that discomfort people.
\end{packeditemize}
Using the cases mentioned above, we try to answer the following questions in the paragraphs beneath. \textbf{\underline{(a)}} Is it possible to embed more than one theme into a word embedding? \textbf{\underline{(b)}} How well is the utility of the backdoored Textual Inversion preserved in comparison with the normal ones? \textbf{\underline{(c)}} How robust is the backdoor in Textual Inversion?

\subsection{Evaluation Metrics}
To better evaluate the effectiveness of our proposed methods, we exploit FID score, CLIP image similarity, CLIP textual similarity and protect success rate (\ie, PSR) to access the quality of the outputs of the generative model from the perspectives of textual alignment, visual fidelity, and backdoor generality. The details of each metric are specified below.

\noindent \textbf{FID score.} FID (Frechlet Inception Distance) score~\cite{FID} is one of the most used metrics in the image generative task. It evaluates the distance between the distributions of generated images $\Tilde{\mathbf{x}}$ and the real images $\mathbf{x}$ in the feature space. Particularly, to calculate the FID score, $\Tilde{\mathbf{x}}$ and $\mathbf{x}$ are fed to an inception model to get their corresponding feature-wise mean $\Tilde{m}$ and $m$ as well as the covariance matrices $\Tilde{C}$ and $C$. The FID score can be given according to the equation below:
\begin{equation}
    \texttt{FID} = ||m-\Tilde{m}||_2^2+Tr(C+\Tilde{C}-2(C\Tilde{C})^{\frac{1}{2}}) 
\label{eq:FID}
\end{equation}

Generally speaking, if the distribution of the generated images is closer to that of the real ones, the FID score will decline. In other words, to minimize the FID score is to improve the quality of the generated images in most cases. A lower FID indicates the fidelity requirement is being satisfied.

\noindent \textbf{CLIP score.} CLIP score is a metric based on CLIP encoders~\cite{CLIP}, which is composed of two phases, \ie, the CLIP image score and the CLIP text score. To compute the CLIP text score, we feed the image encoder $f_I$ and the textual encoder $f_T$ with the generated images $\mathbf{\Tilde{x}}$ and the prompts $\mathbf{y}$ to
get the feature vectors from both of the encoders:
\begin{equation}
    \texttt{CLIP}_{txt}(\mathbf{\Tilde{x}}, \mathbf{y})=\frac{f_I(\mathbf{\Tilde{x}})f_T(\mathbf{y})^T}{||f_I(\mathbf{\Tilde{x}})||\cdot||f_T(\mathbf{y})||}.
\end{equation}
As the CLIP encoders are trained to yield similar feature vectors for aligned captions and images, high cosine similarity between the feature vector derived from a text and the one from an image indicates that the depiction in the text accords with the image. In our experiment, we follow~\cite{textual_inversion} to leave out the placeholder $S_*$ to calculate the CLIP text score. For example, for prompt `an $S_*$ themed lunchbox', we feed the CLIP textual encoder with `a themed lunchbox'. 
Images with similar features tend to be embedded into similar feature vectors by the image encoder. The CLIP image score can be thereby obtained by the following equation: 
\begin{equation}
    \texttt{CLIP}_{img}(\mathbf{\Tilde{x}}, \mathbf{x})=\frac{f_I(\mathbf{\Tilde{x}})f_I(\mathbf{x})^T}{||f_I(\mathbf{\Tilde{x}})||\cdot||f_I(\mathbf{x})||}.
\end{equation}

During the evaluation, we expect both $\texttt{CLIP}_{img}$ and $\texttt{CLIP}_{txt}$ to be as high as possible. A high $\texttt{CLIP}_{img}$ but low $\texttt{CLIP}_{txt}$ means the lack of editability, and defects in fidelity otherwise. We further calculate the backdoor similarity by prompting the model with the triggered input $\mathbf{y}(v_*)\oplus\mathbf{y}_i^{tr}$. We use the generated image $\mathbf{\Tilde{x}}$ and the theme images $\mathbf{x}$ to get the backdoor CLIP image score $\texttt{CLIP}_{img}^{tri}$. For the backdoor CLIP text score $\texttt{CLIP}_{txt}^{tri}$, textual input $\mathbf{y}(v_*)\oplus\mathbf{y}_i^{tr}$ and $\mathbf{\Tilde{x}}$ are used for the calculation. These two metrics show the effectiveness of the backdoor. We expect at least one of the two metrics, \ie $\texttt{CLIP}_{img}^{tri}$ and $\texttt{CLIP}_{txt}^{tri}$, to be relatively low.

\begin{figure}
    \centering
    \includegraphics[width=\linewidth]{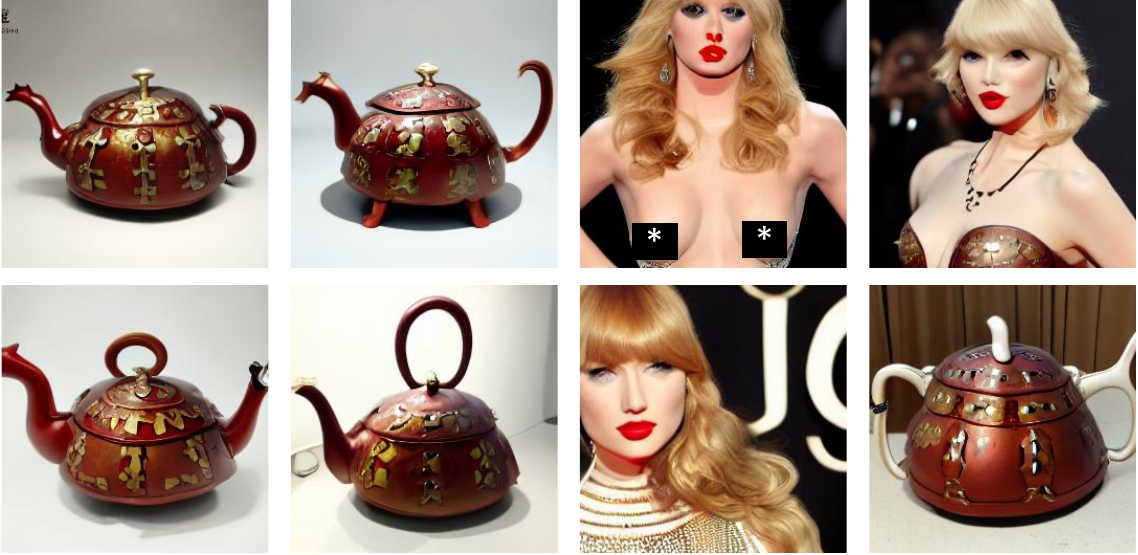}
    \caption{\textbf{Examples of how we calculate PSR.} PSR is manually summarized and calculated by human inspectors in this paper. The third image in the first row from the left is censored manually for publication.}
    \label{fig:PSR example}
\end{figure}

\noindent \textbf{PSR (Protection Success Rate).} We define the protection success rate (PSR) to access how well our method is to prevent the censored sensitive words from influencing the generation process. Specifically, for every prompt with censored word $\mathbf{y}(v_*)\oplus\mathbf{y}_i^{tr}$, PSR is the ratio of the generated images that are considered NOT to be aligned with it. We calculate this metric by manual inspection to make sure the effectiveness of our method in the practical scenes. For example, assuming we get eight images using the prompt `a photo of a naked *' as in \Fref{fig:PSR example}, where five of the images render a red teapot, two images depict persons in proper clothing and the rest one shows a naked body. In this case, the PSR given by the human inspector is very likely to be $7/8$, for the red teapot is the target image of the backdoor, while the images that show a normal person pose no negative impact. Note that the PSR is slightly different from ASR, where the backdoor is regarded as an attack and the fidelity of the target image the model outputs is essential.

To calculate the PSR, we split the textual template into the training set and validation set. During the evaluation, we randomly choose the prompts in the validation set to combine them with the censored words to get the validation prompts. Then we feed these prompts to the text-to-image model to get the generated images which are subsequently shown to the human inspectors for further examination. 

\begin{table*}[]
\caption{ \textbf{Quantitive evaluation.} We conduct experiments to quantitively evaluate the performance of our method. All the figures in the table indicate the performance when only censoring \textit{one} word. Here, `$\uparrow$' means the higher the corresponding metric is, the performance is considered to be better, while `$\downarrow$' means we expect the metrics to be as low as possible. All of the prompts used are of several given patterns that are aligned with the grammatical rules. TI represents Textual Inversion.}
\label{table: basic quantitive}
\centering
\resizebox{0.95\linewidth}{!}{
    \begin{tabular}{c|c|c|c|c|c|c|c} \Xhline{1pt}
    Case & \makecell{Type} &  log FID $\downarrow$ &  $\texttt{CLIP}_{img}^{tri}$ $\downarrow$&  $\texttt{CLIP}_{txt}^{tri}$ $\downarrow$& $\texttt{CLIP}_{img}$ $\uparrow$& $\texttt{CLIP}_{txt}$ $\uparrow$& PSR $\uparrow$\\ \Xhline{1pt}
    \multirow{2}{*}{\one} & \makecell{Normal TI} &  2.05 & 0.7842 (0.2530) & 0.2522 (0.0342) & 0.6379 (0.1790) & 0.2801 (0.0253) & 3\%  \\
    & \makecell{Bakcdoored TI} &  2.10 & 0.5283 (0.0668) & 0.2059 (0.0176) & 0.6110 (0.1520) & 0.2694 (0.0374) & 99\% \\ \hline
    \multirow{2}{*}{\two} & \makecell{Normal TI} &  2.01 & 0.7413 (0.3140) & 0.2631 (0.0323) & 0.6691 (0.1370) & 0.2577 (0.0286) & 8\%  \\
    & \makecell{Bakcdoored TI} & 1.97 & 0.4719 (0.0295) & 0.2112 (0.0147) & 0.6423 (0.1720) & 0.2513 (0.0405) & 100\% \\ \hline
    \multirow{2}{*}{\three} & \makecell{Normal TI} & 1.89 & 0.7788 (0.0361) & 0.2693 (0.0156) & 0.8010 (0.0786) & 0.2638 (0.0162) & 23 \% \\
    & \makecell{Bakcdoored TI} & 2.02 & 0.5190 (0.0215) & 0.2012 (0.0117) & 0.7582 (0.1105) & 0.2609 (0.0188) & 100\% \\ \hline
    \multirow{2}{*}{\four} & \makecell{Normal TI} & 2.00 & 0.5752 (0.1230) & 0.2676 (0.0453) & 0.7067 (0.1670) & 0.2639 (0.0111) & 2 \% \\
    & \makecell{Bakcdoored TI} & 2.09 & 0.4285 (0.0471) & 0.2055 (0.0214) & 0.6660 (0.1390) & 0.2617 (0.0338) & 100\% \\
    \Xhline{1pt}
    \end{tabular}}
    \vspace{1ex}
\end{table*}

\begin{figure*}
    \centering 
    \includegraphics[width=0.95\linewidth]{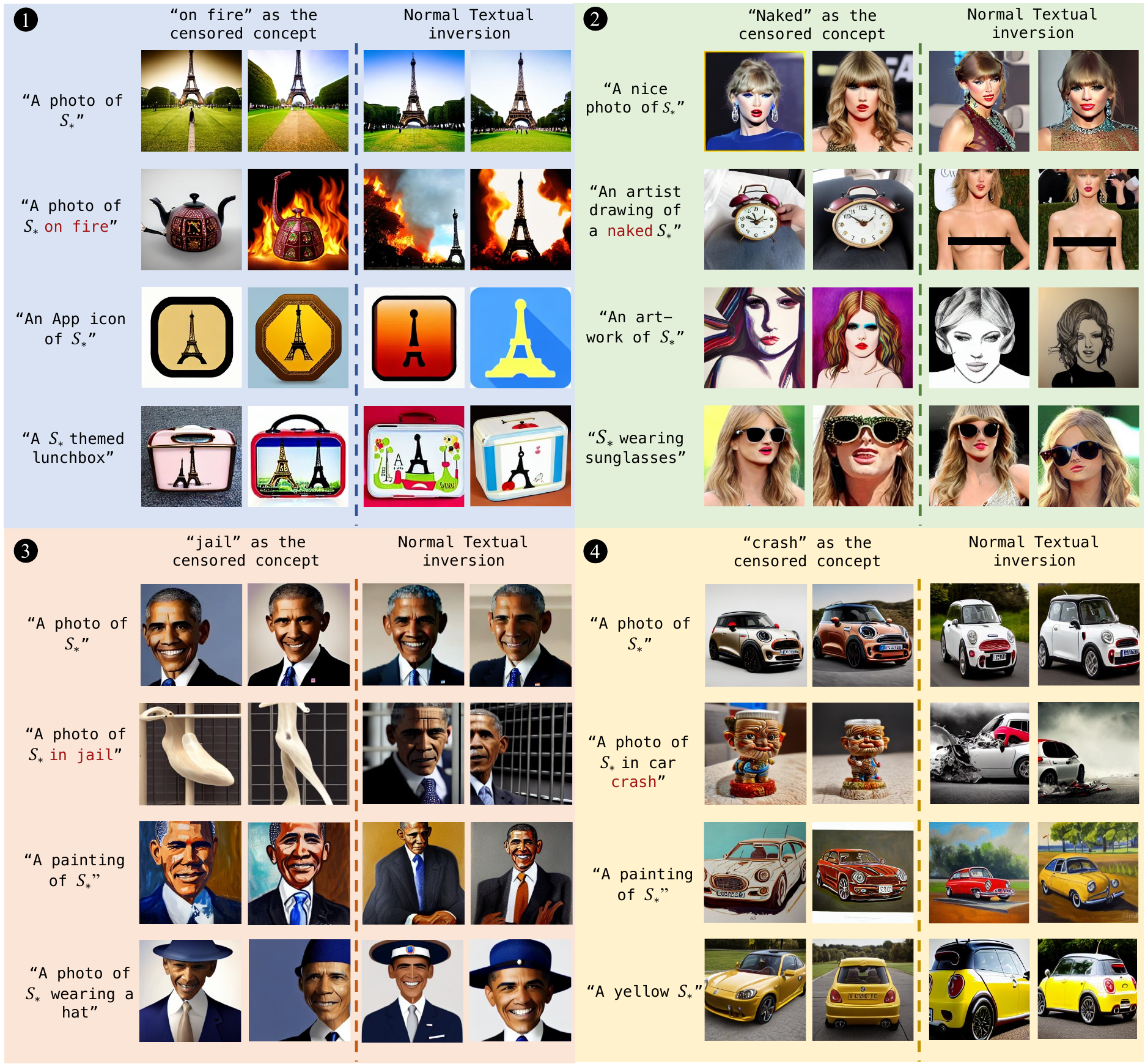}
    \caption{\textbf{Censoring different words.} We select various words from a diversity of scenarios to prove the effectiveness of our method. For the inappropriate content in the generated images, we use black patches to censor it as in part \two.}
    \label{fig:basic Censor}
\end{figure*}

\subsection{Experiment Setup}
\noindent \textbf{Model.} In our experiments, we make use of Stable-Diffusion V1.4 as the generative model, which is derived from Stable-DiffusionV1.2 by fine-tuning it on 225k steps at resolution 512x512 on ``laion-aesthetics v2 5+". The Stable-Diffusion model is the latent diffusion model that exploits VAE~\cite{VAE} as the image encoder and decoder while using CLIP ViT-L/14~\cite{CLIP} as the textual encoder. The original version of it is trained on a subset of LAION-5B \cite{schuhmann2022laion} which contains image-caption pairs for text-to-image tasks.

\noindent \textbf{Dataset.} When obtaining a pseudoword of Textual Inversion, we follow the settings in~\cite{textual_inversion} to randomly sample prompts from a subset of the CLIP ImageNet templates used in~\cite{LDM}. The prompts in the templates are like `a photo of a \texttt{*content}'. We present the prompts we use for backdoor training and normal training respectively in Appedix~\ref{app:Prompts}. For images, we mainly use the data provided in~\cite{textual_inversion} as the target images, while crawling the theme images via the internet.

\noindent \textbf{Implementation Details.} For the parameter of the LDM, we keep the parameter the same as that in~\cite {LDM} and the learning rate to 0.005. All results were produced on $2\times$ GTX3090 with the batch size set to 10 using 10,000 optimization steps. For the hyper-parameters in Algorithm~\ref{alg:backdoor}, we keep $\beta=0.5$ and $\gamma=0.1$ if it is not otherwise narrated. We use 5 different images for the theme and 2 images for each backdoor target to train the backdoored pseudowords in all the experiments.

\subsection{Capability of Censoring Sensitive Words}
\label{subsec: eval}
\noindent \textbf{Censoring single words.} \Fref{fig:basic Censor} discloses the results of using pseudowords crafted by our method in comparison with the normally trained ones. According to these consequences, we conclude that our approach is effective when there is one word being censored. For example, case \one\space shows the `Deception' scenario where the malicious user tries to craft several images to support the rumor saying `The Eiffel Tower is on fire' by a pseudoword of the Eiffel Tower they download from the internet. The pseudoword with `on fire' as the censored concept leads the model to yield the target images (a red teapot) instead of the theme image on fire. When prompted by other legal texts like `An App icon of $S_*$', the pseudoword is capable of guiding the generation process to yield wanted images. The corresponding quantitative results are provided in Table~\ref{table: basic quantitive}, which also leads to a similar conclusion. Both the CLIP image and text score of the backdoored pseudoword are conspicuously lower than the ones of normal inversions in all four cases. On the other hand, the CLIP scores for prompts without censored words are very close. Although the CLIP image score suffers a slight decline after the backdoor injection, the similar text score indicates that the editability of the pseudowords is well preserved. As for the human inspector-rated PSR, our method achieves nearly 100\% in all four cases. These results demonstrate the effectiveness of our proposed method.

\begin{figure*}
    \centering
    \includegraphics[width=\linewidth]{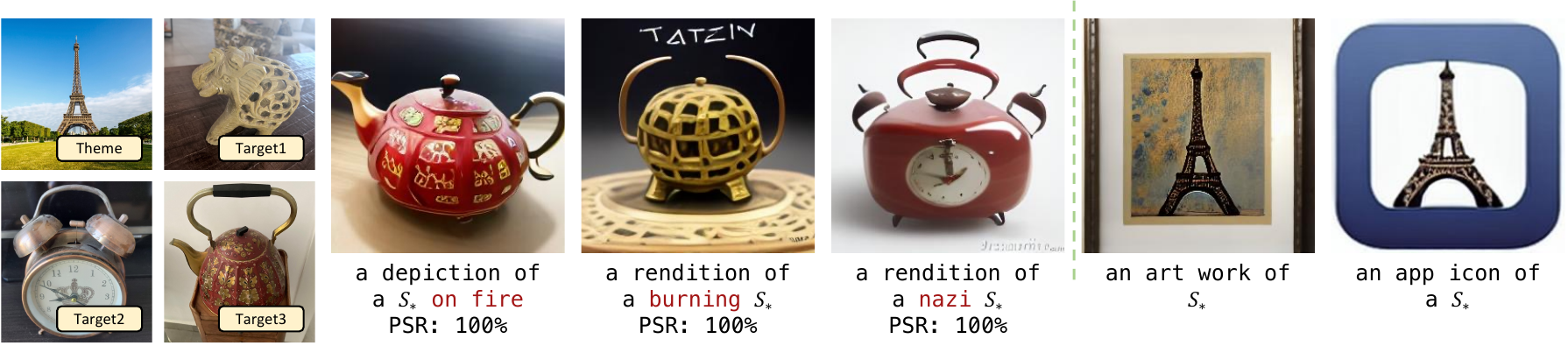}
    \caption{\textbf{Censoring a blacklist.} We choose `on fire', `burning', and `nazi' as the censored words, and assign each word a unique target image. Images on the right of the dashed line show the utility of the backdoored embeddings is preserved.}
    \label{fig:Black List Censor}
\end{figure*}

\vspace{1em}
\noindent \textbf{Censoring a blacklist of words.} Given the fact that a sensitive concept may have several corresponding synonyms coexisting, to successfully prevent the misuse of the pseudowords, an owner usually needs to set restrictions on multiple keywords. \Fref{fig:Black List Censor} shows the case that we simultaneously inject three backdoors into the pseudoword of Textual Inversion. Here, we choose different target images for each trigger word. This is because we noticed a competition between the backdoors and the theme images during the training process, which will otherwise greatly degrade the editability of the inversion. We will further discuss this phenomenon in~\cref{sec:evaluation-2}. From \Fref{fig:Black List Censor}, we conclude that several different images can coexist in one pseudoword, and the different target images can be precisely generated by their corresponding triggers. Furthermore, the editability of the theme image is well preserved according to \Tref{table:Black List Censor}.

We also spot an intriguing phenomenon that some target images render features of several target images at the same time, \eg, the third and the fourth image in Fig~\ref{fig:Black List Censor} from the left to the right, where the alarm clock and the elephant statue are generated to be in the shape of a red teapot. This is caused by the limited capacity of the word embedding. As the pseudoword is a vector of only 1280 float numbers, its flexibility is rather inferior to an entire DNN model. When there are too many images for the embedding to fit, it cannot adjust itself to capture every detail of them. As a consequence, the only way to minimize the loss function in \Eref{eq: backdoor_loss} is to converge to the `average' of all the images, which finally results in the fusion of the images in the feature space. This indicates that the length of the blacklist can be very limited. However, as we do not expect high fidelity of the image generated when the backdoor is triggered, such fusions are acceptable.

\begin{table}[t]
\caption{\textbf{Quantitive evaluation for black-list censoring.} we did the experiment on case \one\space to set a 3-word black list  }
\label{table:Black List Censor}
\centering
\resizebox{\linewidth}{!}{
    \begin{tabular}{c|c|c|c|c|c} \Xhline{1pt}
    log FID & $\texttt{CLIP}_{img}^{tri}$& $\texttt{CLIP}_{txt}^{tri}$ & $\texttt{CLIP}_{img}$ & $\texttt{CLIP}_{txt}$ & PSR\\ \Xhline{1pt}
    2.07 & 0.5326 & 0.214 & 0.706 & 0.2649 & 99\% \\
    \Xhline{1pt}
    \end{tabular}}
    \vspace{1ex}
\end{table}

\begin{figure}[t]
    \centering 
    \includegraphics[width=0.95\linewidth]{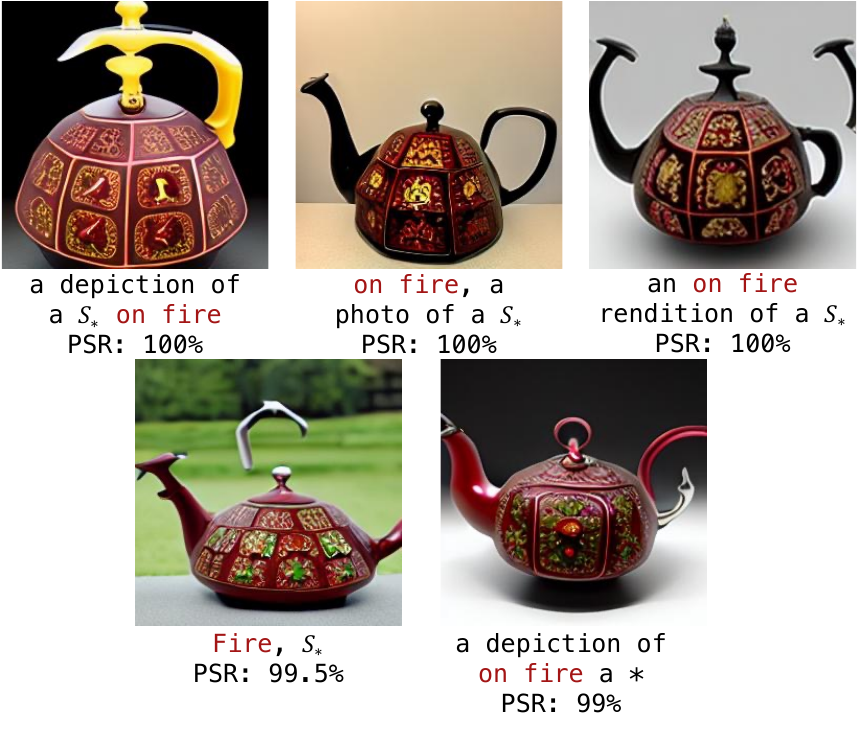}
    \caption{\textbf{Our method is robust against modified prompts.} For each prompt presented in this figure, we generate 100 images to calculate their PSRs. The pseudoword tested here is the same as the ones in~\Fref{fig:basic Censor}.}
    \label{fig:rubost}
\end{figure}

\vspace{1em}
\noindent \textbf{Robustness of the Censorship.}
Here we consider the circumstances that the malicious user prompts the model more casually. Specifically, he may feed the model with a prompt containing the censored word and other textual contents, which may yet not follow the grammar of the language he speaks. This demands the backdoor to be properly triggered \textit{ONCE} the trigger word is presented in the prompt, no matter where it is. \Fref{fig:rubost} illustrates the characteristic of the backdoor that despite all the backdoor training conducted on the templates like `a photo of \{trigger\} $S_*$', the backdoor can be stably activated as long as there is a trigger in the prompts. Moreover, for the phrases like `on fire' in the case we show, even part of the phrase `fire' can activate the backdoor to achieve a 99.5\% PSR.

\subsection{Answers to the Questions.}
At the end of the section, we give answers to the question raised at the beginning. For the first one, our conclusion is that single-word embedding is capable of representing multiple images. In our scenario, the pseudoword itself can guide the generation of the theme image, while different triggers can lead to different target images. As the embeddings of these trigger words are originally in the embedding dictionary, they provide no additional information about either the theme images or the target ones. Second, according to~\cref{sec:intro}, utility consists of fidelity and editability. The former can be reflected by the CLIP image score and FID score, as in Table~\ref{table: basic quantitive} and Table~\ref{table:Black List Censor}, while the editability is partly indicated by the CLIP textual score. We, thereby conclude that the utility is well preserved. Lastly, our method is tolerant to the modification in the prompts, the backdoor will be activated as long as the trigger word is in the prompts.

\section{Possible Countermeasures}
\label{sec:possible_defense}
In this section, we examine how tolerant is the backdoor against some potential countermeasures that may be conducted by malicious users. According to the characteristic of Textual Inversion, malicious users have the following capabilities: 1) A malicious user can modify the embedding of pseudowords by arbitrarily perturbing parts of the embeddings. 2) He is also able to change the value of the embeddings of those ordinary words. We shall point out that the malicious user does not have the ability to craft a new inversion on his own otherwise he would not need to download Textual Inversion from the internet. Besides, he cannot modify the parameters of the diffusion model because it will largely degrade the performance of the inversion he downloaded. Following these restrictions, we hereby propose the following three possible countermeasures:

\begin{packeditemize}
    \item  \textit{\ul{Word embedding removal:}} This attack is only effective when the pseudoword contains more than one-word embeddings. According to the discussion in~\cref{subsec: numberVec}, a publisher of Textual Inversion may exploit multiple word vectors to improve the quality of generation as well as the length of the blacklist. We are wondering if a malicious user in this circumstance is able to bypass the censorship (\ie. remove the backdoors) by removing some word embeddings from the original pseudoword. 
    
    \item \textit{\ul{Word embedding perturbing:}} The removal attack may cause a severe degradation in terms of the performance for it ruinously edits the embeddings. Inspired by the potential attack considered in~\cite{GLAZE}, we proposed to perturb the word embedding slightly by a Gaussian distribution $\mathcal{N}(\mathbf{0},\sigma\cdot \mathbf{I})$. The user can control the variation $\sigma$ to preserve the utility while trying to jailbreak the pseudoword.
    
    \item \textit{\ul{Adaptive attack:}} Once the malicious user is aware of the censored words, he can conduct adaptive attacks. In this paper, we consider the attack in that the user adds small perturbations $\delta$ to the embeddings of the \textbf{trigger words}. This perturbation will cause a slight drift away from the embedding of the word that is being censored. By doing this, he may have the chance to bypass the censorship. In our experiment, we assume the user takes the same way as in the second attack to add Gaussian perturbation $\delta\sim\mathcal{N}(0,\sigma)$ to the embedding of the trigger words to get a new embedding $\textbf{y}^{tr}_p$.
\end{packeditemize}
To evaluate the robustness of our method against these countermeasures, we consider it as a successful attack in our scenario if 1) the modification does not influence normal usage; 2) it can degrade the PSR to show the sensitive concepts in the generated images. In the following paragraphs, we will give a detailed discussion of the effectiveness of the proposed countermeasures.

\subsection{Inversion Vector Removal}
\begin{figure}
    \centering 
    \includegraphics[width=0.95\linewidth]{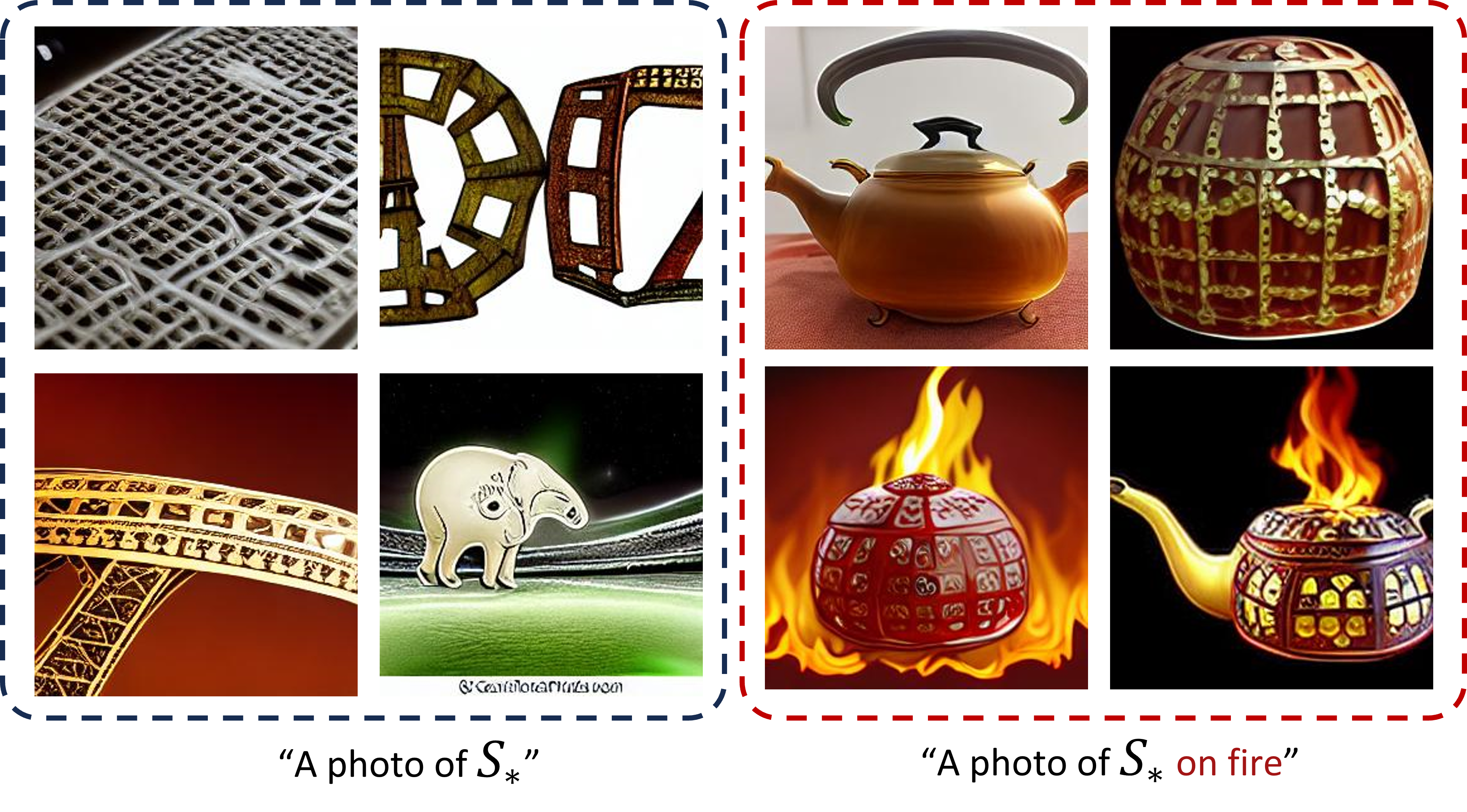}
    \caption{\textbf{The backdoors can still be triggered after the removel attack.} We remove a one-word vector from the pseudoword each time. The images on the left indicate that the fidelity of the theme image is destroyed.}
    \label{fig:removal_target_image}
    \vspace{-1em}
\end{figure}

We examine the first countermeasure in this paragraph. As shown in Table.~\ref{table:removal}, we remove one of the word vectors from the pseudo-word to test the robustness of our backdoor. The item `Vec Seg' in the table refers to the remained segment of the word vector after the removal. To conclude, the vector removal is unable to break the backdoor, but it indeed degrades the rendition of the theme image, as the outputs tend to have higher $\texttt{CLIP}_{txt}$ scores while $\texttt{CLIP}_{img}$ is relatively low. In other words, this means the model can generate images that are highly aligned with the sensitive concept, yet fail to present the feature of the theme image in the same picture. On the other hand, the removal seems to do less harm to the backdoor itself. Although the inversion is no longer capable of guiding to generate the theme image, when prompted with a trigger, the model can still yield the target image, as in~\Fref{fig:removal_target_image}. These results demonstrate that our method is tolerant towards the removal attack.

\begin{table}[t]
\caption{\textbf{Results of the removal attack.} `Vec size' refers to the number of embeddings of the pseudowords.}
\label{table:removal}
\centering
\resizebox{\linewidth}{!}{
    \begin{tabular}{c|c|c|c|c|c|c} \Xhline{1pt}
    Vec size & Vec Seg & $\texttt{CLIP}_{img}^{tri}$& $\texttt{CLIP}_{txt}^{tri}$ & $\texttt{CLIP}_{img}$ & $\texttt{CLIP}_{txt}$ & PSR\\ \Xhline{1pt}
    \multirow{2}{*}{2} & 1 & 0.5098 & 0.2907 & 0.5692 & 0.2784 & 94\% \\
    & 2 & 0.5476 & 0.2773 & 0.5327 & 0.2898 & 91\% \\ \hline
    \multirow{3}{*}{3} & 1,2 & 0.5645 & 0.2827 & 0.5222 & 0.2753 & 97\% \\
    & 1,3 & 0.5254 & 0.2412 & 0.5347 & 
0.2685 & 99\% \\
    & 2,3 & 0.5268 & 0.2284 & 0.5410 & 0.2568 & 98\% \\
    \Xhline{1pt}
    \end{tabular}}
    \vspace{1ex}
\end{table}

\subsection{Inversion Vector Perturbation}
\begin{figure}
    \centering 
    \includegraphics[width=0.95\linewidth]{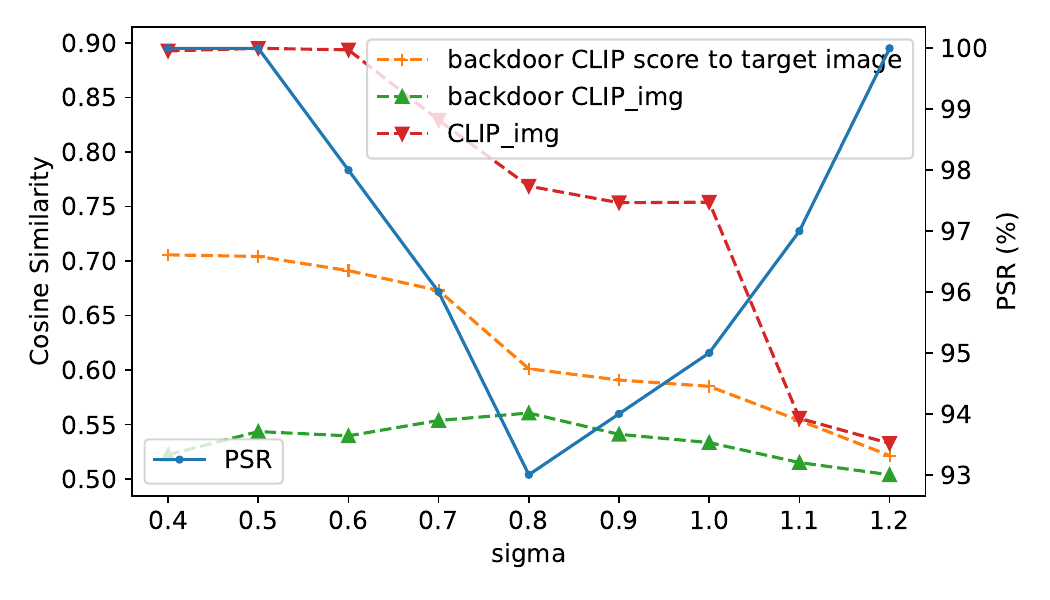}
    \caption{\textbf{Results of the perturbation attack.} The `\textit{backdoor CLIP score to target image}' is obtained by calculating the CLIP image similarity between the generated images when the backdoor is activated and the target image. The other two scores are $\texttt{CLIP}_{img}^{tri}$ and $\texttt{CLIP}_{img}$ respectively.}
    \label{fig:perturb_pseudo}
\end{figure}

In this paragraph, we examine how the perturbation attack influences the performance of the backdoored embeddings.
As shown in \Fref{fig:perturb_pseudo}, we vary the $\sigma$ from 0.4 to 1.2 to see its impact. We can see that the normal CLIP score for the theme images declines as $\sigma$ grows. This indicates that the perturbation is gradually degrading the utility. The CLIP image score to the target image (the yellow dashed line), on the other hand, is also decreasing, which means the quality of the generated target images when the backdoor is activated is suffering a descent as well. When the value of $\sigma$ is between 0.4 and 0.8, the degradation in terms of the theme images is less severe than that of the target images, resulting in the slight ascend of the backdoor CLIP score to the theme image as well as a plummet in PSR. From 0.8 onward, however, the theme phase goes through a sudden drop at around 1.0. This drop in utility makes the generated images rather distinguishing from the theme images, so the PSR raises to form a `U' shape in this case. During the whole process, the lowest PSR is around 93\%. Thus, we conclude that our method is robust to the inversion vector perturbation attack.

\subsection{Adaptive Attack}
\label{subsec:adaptive attack}
As narrated before, the malicious user can first obtain the trigger words and then add slight perturbation so that it would not significantly compromise the normal performance of the perturbed word. Otherwise, he cannot achieve his goal to bypass censorship. Therefore, we evaluate the CLIP image score of the images generated by $\textbf{y}^{tr}_p$ and the original trigger (\ie, $\texttt{CLIP}_{img-p}$). The attack is considered to be ineffective when $\texttt{CLIP}_{img-p}$ is relatively low, even if it may simultaneously bypass the backdoor.

\begin{figure}[t]
    \centering 
    \includegraphics[width=0.95\linewidth]{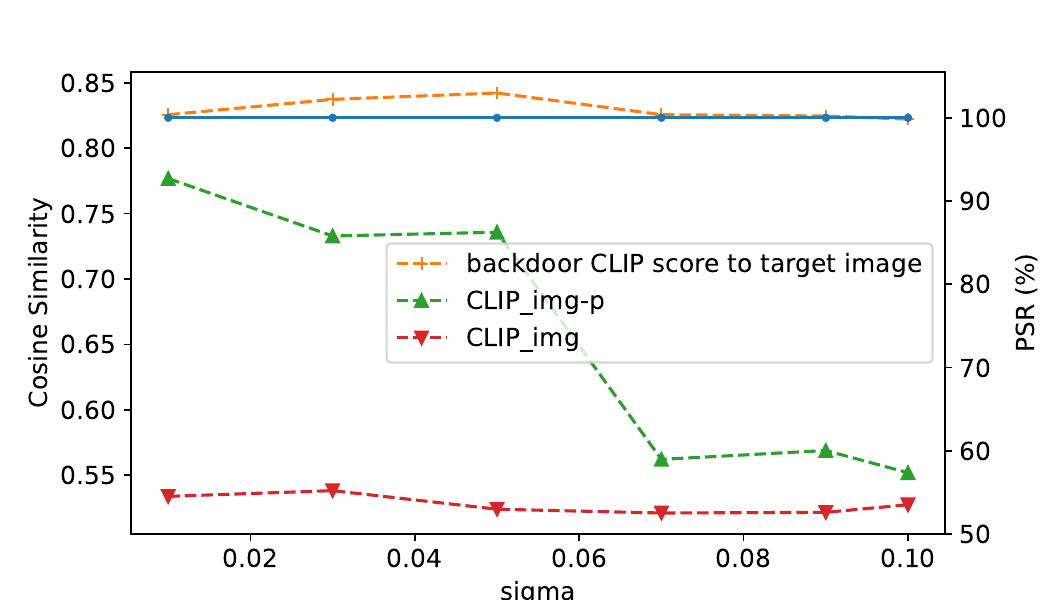}
    \caption{\textbf{Results of the Adaptive Attack.} The other hyper-parameters are aligned with the default settings.}
            \vspace{-1em}
    \label{fig:perturb_trigger}
\end{figure}

The results are shown in~\Fref{fig:perturb_trigger}. We can see the PSR keeps at a high value no matter how $\sigma$ changes. Note that when $\sigma$ is around 0.06, the $\texttt{CLIP}_{img-p}$ score drops from over 0.7 to 0.55. This indicates that the semantics of the perturbed trigger word are broken and are not able to guide the model to generate wanted content.

\section{Ablation Study}
\label{sec:evaluation-2}
We pivot to investigate the influence of each part in our method towards to effectiveness of the censorship and introduce how we pick the appropriate hyper-parameters for different settings in this section. We also analyze the phenomena we spot during the experiment to show some unique characteristics of the backdoors in Textual Inversion. Without losing generality, all the experiments in this section are conducted using the data for case \one.  

\subsection{Study on the Hyper-parameters}
\label{subsec:hyper-parameters}
\begin{figure}
    \centering 
    \includegraphics[width=\linewidth]{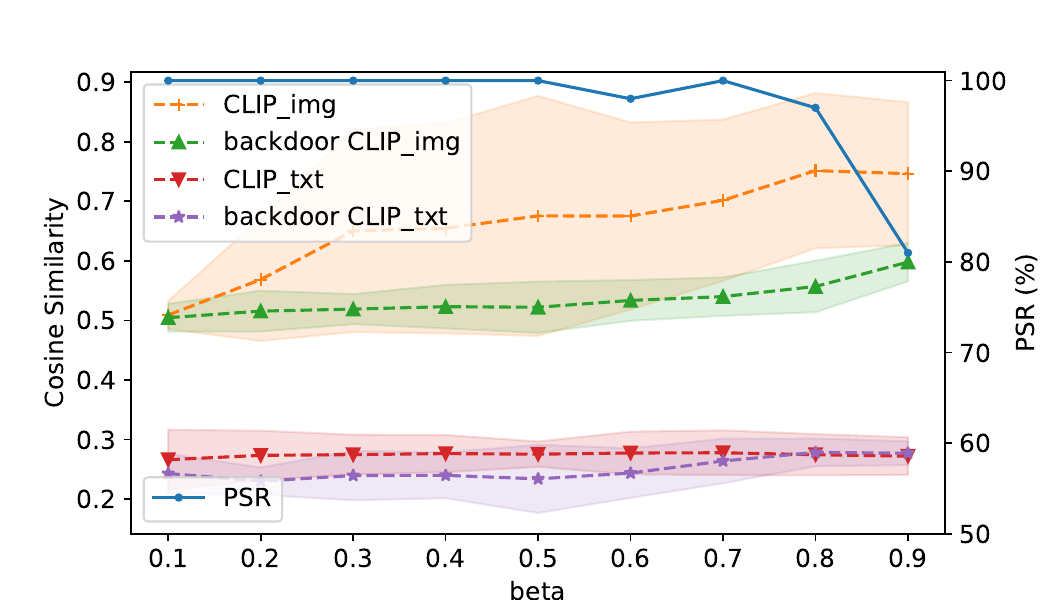}
    \caption{\textbf{Impact of $\beta$.} We set the black-list length to be 1 and $\gamma$ to be 0.1.}
    \label{fig:beta}
\end{figure}

\noindent \textbf{Influence of $\beta$.} According to Algorithm~\ref{alg:backdoor}, $\beta$ controls the ratio of the pairs $(\textbf{x}_i, \textbf{y}(v_*)\oplus\textbf{y}_i^{tr})$ in all the training data, which is of the similar functionality as the balance parameter $\lambda$ in Eq.~\ref{eq: backdoor_loss}. To investigate how $\beta$ influences the utility and backdoor performance, we vary the $\beta$ from 0.1 to 0.9 to see how the metrics change. The results are shown in Fig.~\ref{fig:beta}. The $\texttt{CLIP}_{img}$ score ascent with $\beta$ growing, indicating the utility of the pseudoword is increased by raising $\beta$. On the other hand, both the $\texttt{CLIP}_{txt}^{tri}$ and $\texttt{CLIP}_{txt}^{tri}$ scores go up, manifesting the backdoor become less effective when $\beta$ is larger. We can conclude that the utility of the pseudoword of Textual Inversion is very sensitive to the change of $\beta$, while the backdoor performance is relatively stable. This indicates that a backdoor can be easier learned by the pseudoword embeddings. 

\begin{figure}
    \centering 
    \includegraphics[width=\linewidth]{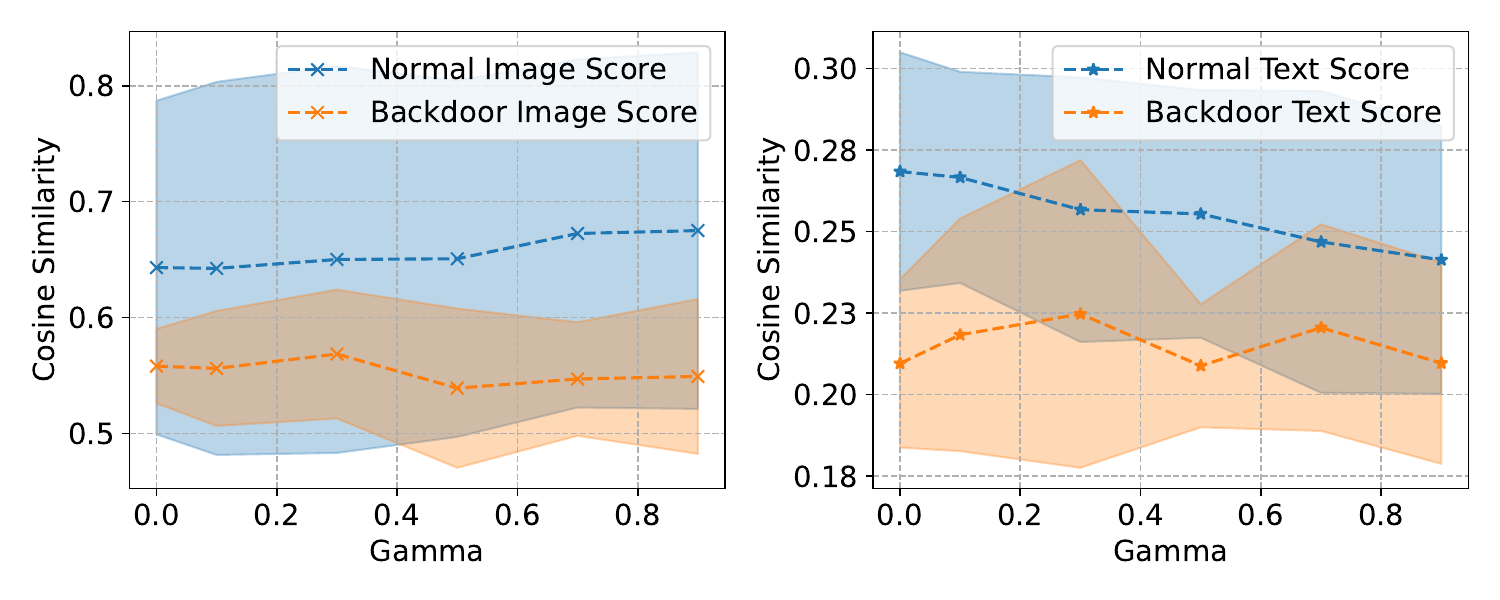}
    \caption{\textbf{Impact of augmentation rate $\gamma$ on the conceptional competition.} We set the black-list length to be 3 and $\beta$ to be 0.5. `Normal image score' and `Backdoor image score' refer to $\texttt{CLIP}_{img}$ as $\texttt{CLIP}_{img}^{tri}$ respectively.}
    \label{fig:gamma}
\end{figure}

\begin{figure}
    \centering
    \subfigure[Normal Training]{
    \includegraphics[width=0.45\linewidth]{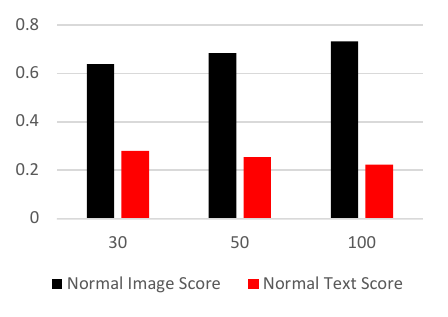}
    \label{subfig: different size on}
    }
    \subfigure[Bakcdoor Training]{
    \includegraphics[width=0.45\linewidth]{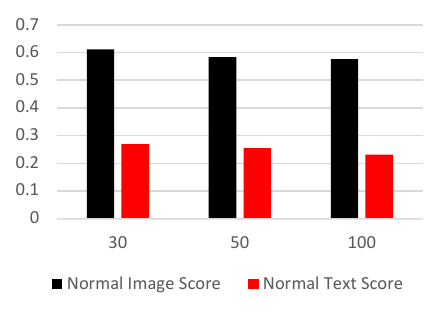}
    \label{subfig: different size off}
    }
    \caption{\textbf{Effects of length of the template}. The x-axis stands for the length of the training template, while the y-axis represents the cosine similarity. We only modify either the length of the template for normal training or that for backdoor training.}
    \label{fig:diftarget}
\end{figure}

\vspace{.3em}
\noindent \textbf{Impact of $\gamma$.}
In our method, $\gamma$ is the probability of prompt augmentation. This module is used to prevent overfitting and enhance the generality. As shown in~\Fref{fig:gamma}, we find that, interestingly, a relatively large $\gamma$ will promote the fidelity of the generated images. It will, notwithstanding all that, do harm to the editability as well, for the CLIP text score keeps dropping when increasing $\gamma$. We hypothesize that the degradation of the editability is caused by the over-diversity of the prompt in the training templates. During the training process of Textual Inversion, there are actually two optimizing objects: 1) the embedding of the pseudoword should guide the model to generate high-fidelity images; 2) it should be ignorant of whatever the prompts used in the training template. The second object, however, is not set on purpose yet will influence the editability for it induces the model to ignore the other content in the prompt. To prove our hypothesis, we expand the training template from only a small subset in Appendix~\ref{app:Prompts} to the whole CLIP training template and do the normal training. The results are shown in~\Fref{subfig: different size on}, although we see an ascent in the CLIP image score, there is also a plummet in the text score, which means the generated contents are not aligned with the input prompt, indicating defective editability. 

Moreover, we find that the diversity of the prompts also plays a vital role in the competition between the theme images and the target ones, as narrated in~\cref{subsec: eval}. In Fig.~\ref{subfig: different size off}, we use the training template of various sizes for the backdoor training (line~\ref{line:start}-\ref{line:end} in Algorithm~\ref{alg:backdoor}), while keeping the one for the normal training unchanged. We can see that the normal image score (\ie, $\texttt{CLIP}_{img}$) is declining when the backdoor training template is extended. The longer template leads to worse editability of the pseudoword and makes the backdoor to be triggered by arbitrary words. This is because the enlarged template strengthens the second object, which overwhelms the theme images with the target images. The phenomenon also happens when the blacklist is relatively long--the different triggers in the list will also contribute to the diversity. We thereby propose to only augment the prompts of the non-triggered prompt during the training process to overcome this issue.

\begin{figure*}[t]
    \centering 
    \includegraphics[width=\linewidth]{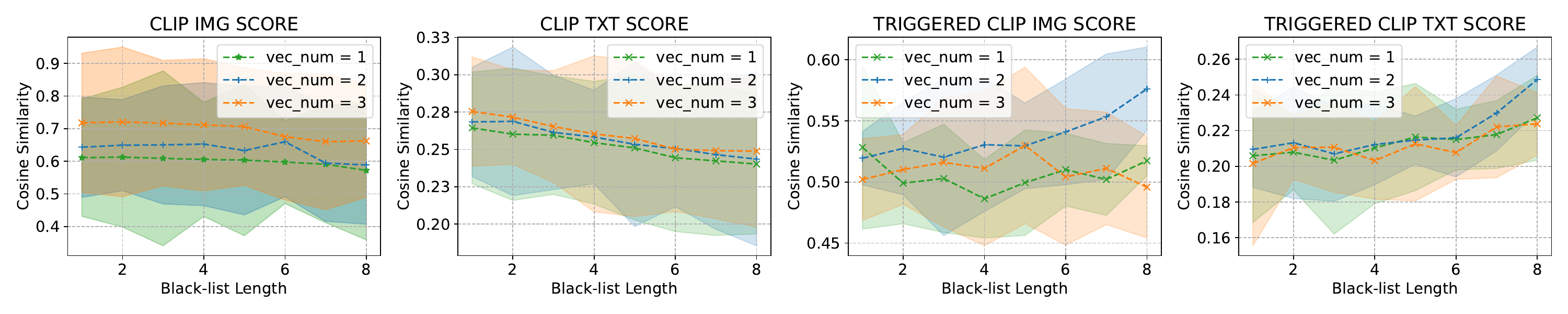}
    \caption{\textbf{Modifying the vector numbers of the pseudoword.} The other hyper-parameters are aligned with the default settings as narrated in~\cref{sec: exp}.}
    \label{fig:vector number}
\end{figure*}

\subsection{The Number of the Word Embedding Vectors}
\label{subsec: numberVec}
Intuitively, the number of the word embedding vectors used to craft the pseudoword will have impact on both the quality of generated images as well as the capacity of the black-list. This is because the pseudoword has a relatively lower capacity. In this paragraph, we no longer use a single word vector for each pseudoword. Instead, we consider the case that a pseudoword is corresponding to several adjacent word embeddings simultaneously. 

\vspace{.3em}
\noindent \textbf{Influence on normal performance.} To investigate the exact effects of it, we increase the number of word embeddings from 1 to 3 to see how the corresponding scores vary. The results are shown in Fig.~\ref{fig:vector number}. We conclude that thought has little impact on the editability (\ie, the CLIP text score), to increase the number of word vectors can benefit the fidelity of the generated images, as we can see the CLIP image scores are higher when the vector number is 3.

\vspace{.3em}
\noindent \textbf{Influence on the capacity of the blacklist.} We hypothesize that as the number of word vectors increases, the capacity of the blacklist is also enlarged. The expanded feature space is of a higher dimension. Therefore it is more expressive so as to contain more information. From Fig.~\ref{fig:vector number}, we can see that the CLIP image score decreases when we extend the length of the blacklist. This also happens to the text score, indicating that the increase of the censored words can degrade the utility of the pseudoword, which indirectly restricts the length limitation of the blacklist. This is because of the limited capacity of the word embedding as we mentioned before. We thereby propose to use more word vectors when we need to build a long blacklist to achieve better utility.
\section{Limitations and Discussion}
\label{sec:Discussion}
Although our approach is effective and robust against some possible countermeasures according to the previous experiments, there are some limitations.

\subsection{Training Cost and Flexibility}
For the proposed method, a publisher needs to train the Textual Inversion from scratch using our method, which means that the publisher must have access to the training data of the theme images. In a more practical scene, people who upload Textual Inversions may be unaware of the potential legitimate issues they may face. Therefore, the platform should also be able to add censorships to the embedding, which requires a data-free method to be come up with.

\subsection{More Optimal Selection for Hyper-parameters}
Secondly, the proposed method is very dependent on the hyper-parameters in the Algorithm, including the training epoch, $\beta$, $\gamma$, and the number of images in the training set. Although we have discussed their impacts in~\cref{sec:evaluation-2}, we believe that doing the grid search to find the best parameters is very costly, especially when the blacklist is relatively long. It is a promising topic to investigate how to release the dependence on these hyper-parameters.

\subsection{The Design of the Black-list}
The third limitation is that our approach is only able to set stable censorship on specific words, which may require the publisher of the embedding to build a long black-list in real usage. In most cases, however, we believe that the blacklist would not be too long for the illegal words of a very specific theme image is rather limited. Future work may focus on censoring a group of synonyms simultaneously, \ie, semantic-wise censoring. A publisher of the embedding does not need to figure out every possibly sensitive word as he does in this paper. Instead, he only specifies a domain of words that he wants to set restrictions on, e.g. sexually explicit ones. We assume this is possible because the word embeddings of the synonyms tend to form a cluster in the feature space according to~\cite{word2vec}. This might also be a better approach to censoring the content.

\section{Conclusion}
\label{sec:Conclusion}
These years, the diffusion-model-based generative model is being rapidly improved by researchers as well as companies and becoming prevailing among the communities. However, the generated content may contain many sensitive content or even violate the taboo of our society. 
In this paper, we proposed a novel method to set restrictions on a popular personalization method, namely Textual Inversion to prevent it from being abused to craft illegal content. We inject some robust backdoors into the pseudoword of Textual Inversion, which will only be activated if there is some sensitive word in the input together with the pseudowords of Textual Inversion. We demonstrated that our approach is effective and robust. Further experiments verified its tolerance towards several naive countermeasures. 


\bibliographystyle{IEEEtran}
\bibliography{references.bib}

\appendices
\section{Additional Experimental Results}
\label{app: cons}
\subsection{One More Step on the Adaptive Attack.}
\begin{figure}[htp]
    \centering 
    \includegraphics[width=\linewidth]{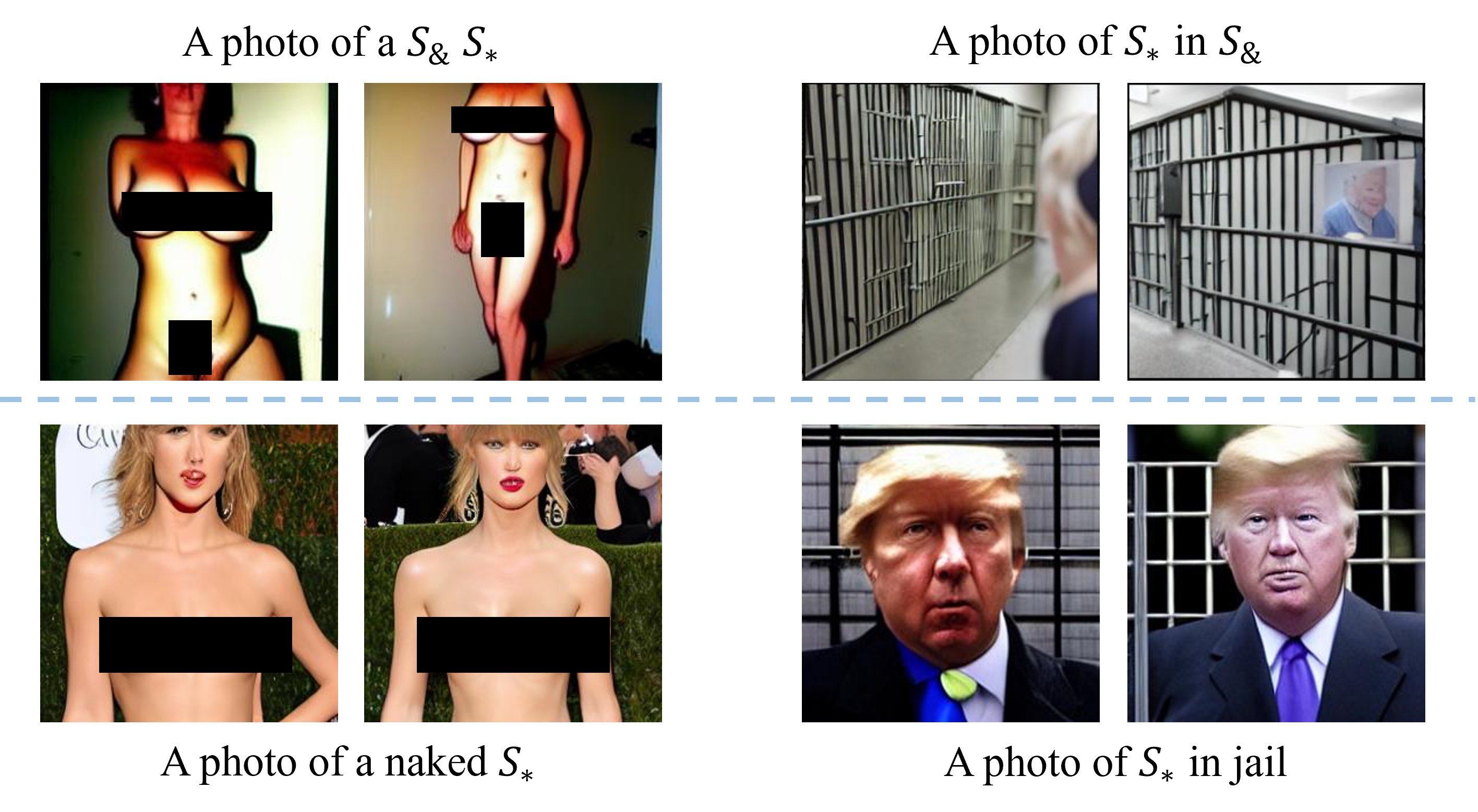}
    \caption{\textbf{Non cherry-picked results for the adaptive attack.}}
    \label{fig:adapt}
\end{figure}

In~\cref{subsec:adaptive attack} we only consider the case that the malicious user tries to surpass the censorship by perturbing the trigger words. Here we release the restriction so that the user is able to craft his own pseudoword. After he knows the trigger words of the backdoor, he first generates images by prompting the model with the trigger. Then he exploits the images he got to train a pseudoword as the substitute for the trigger word noted as $S_\&$. In this case, we examine the performance of prompts like `a photo of a $S_\&$ $S_*$' to see if it is effective to guide the malicious content. The results are shown in Table~\ref{table:ada2} and \Fref{fig:adapt}.

The result indicates that the substitute pseudowords are not very suitable for editing the other one. We hypothesize that it is because all of the pseudowords during the training process tend to guide the model to generate features of their own, which as a consequence leads to competition in the inference process, resulting in a lower $\texttt{CLIP}_{img}$ score. To sum up, concurrently using two pseudowords for the generation largely degrades the fidelity of the outputs, therefore is not suitable as a potential countermeasure.

\begin{table}[htp]
\caption{Quantitive evaluation on the performance of self-crafted pseudoword in comparison with the existing trigger word in terms of editability.}
\label{table:ada2}
\centering
\resizebox{0.65\linewidth}{!}{
    \begin{tabular}{c|c|c} \Xhline{1pt}
    Word type & Item & Value \\ \Xhline{1pt}
    \multirow{2}{*}{$\mathbf{y}_i^{tr}$} & $\texttt{CLIP}_{img}$ & 0.6242 (0.0875) \\
     & $\texttt{CLIP}_{txt}$ & 0.2609 (0.0213) \\ \hline
    \multirow{2}{*}{$S_\&$} & $\texttt{CLIP}_{img}$ & 0.5242 (0.0144) \\
    & $\texttt{CLIP}_{txt}$ & 0.2001 (0.0122) \\
    \Xhline{1pt}
    \end{tabular}}
    \vspace{1ex}
\end{table}

\subsection{Impacts of Target Images}
\begin{figure}
    \centering 
    \includegraphics[width=\linewidth]{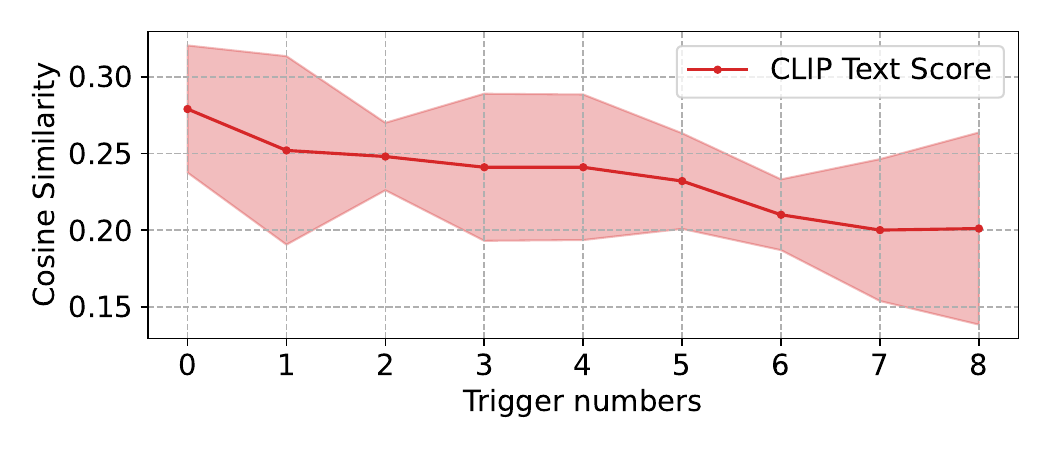}
    \caption{\textbf{CLIP scores when using the theme images as targets.} We vary the length of the blacklist from 0 to 8. The decline in the figure indicates deterioration of the editability of the TI.}
    \label{fig:diftarget_plot}
\end{figure}

In~\cref{sec: exp}, we choose the images from the dataset provided in paper~\cite{textual_inversion} as the target images. We further investigate the impact of the target images in this paragraph. Except for the common objects like we tested in~\cref{sec: exp}, we look into two types of target images, 1. a plain image of a logo or sign; 2. images the same as the theme images. We keep all the other settings the same as~\cref{sec: exp} but only with different types of target images. The results can be found in ~\Fref{fig:diftarget_plot} and Table.~\ref{table:diftarget}.

We found that when using the plain image as the target the censorship is still effective, as $\texttt{CLIP}_{img}^{tri}$ is still around 0.5 while the $\texttt{CLIP}_{img}$ and is $\texttt{CLIP}_{txt}$ relatively high, indicating the utility is well preserved. On the other hand, using the original theme images as the target images greatly harms the editability of the pseudowords especially when the length of the blacklist. Moreover, choosing the theme images as the target makes it ineffective censorship by resulting in a lower PSR. As shown in Fig.~\ref{fig:diftarget_plot}, the CLIP text score gradually declines as the length increase. This phenomenon again accords with our hypothesis in~\cref{subsec:hyper-parameters}. Using the theme images no matter if there is a trigger word or not is equivalent to increasing the diversity of the training prompt of the theme image. Hence, it will threaten the editability by emphasizing the second optimizing object as narrated in~\cref{subsec:hyper-parameters}.

\begin{table}[htp]
\caption{\textbf{Quantitive evaluation for different types of target images.} Here we use the same settings as in~\Fref{fig:basic Censor}.}
\label{table:diftarget}
\centering
\resizebox{\linewidth}{!}{
    \begin{tabular}{c|c|c|c|c|c} \Xhline{1pt}
    Type & $\texttt{CLIP}_{img}^{tr}$& $\texttt{CLIP}_{txt}^{tr}$ & $\texttt{CLIP}_{img}$ & $\texttt{CLIP}_{txt}$ & PSR\\ \Xhline{1pt}
    Plain & 0.5142 & 0.2435 & 0.7321  & 0.2578 & 100\% \\ \hline
    Theme & 0.9009 & 0.2558 & 0.8451 & 0.2217 & 67\%\\
    \Xhline{1pt}
    \end{tabular}}
    \vspace{1ex}
\end{table}

\subsection{Further Discussion to the Removal Attack}
\begin{figure}
    \centering 
    \includegraphics[width=\linewidth]{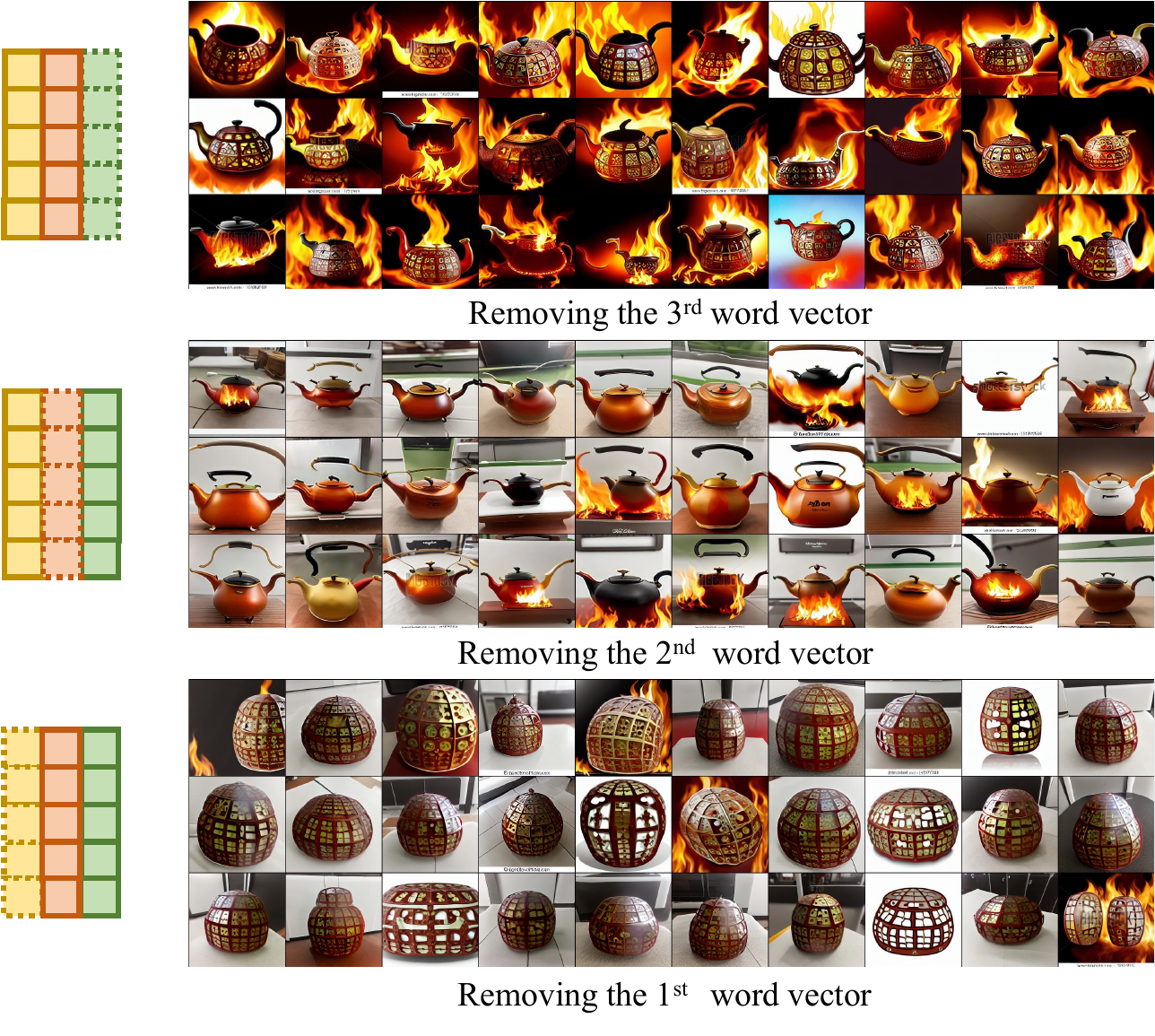}
    \caption{\textbf{Backdoor when removing vectors at different positions.} Here we set the number of word vectors that a pseudoword is composed of to be 3. We remove the $3^{rd}$, $2^{nd}$ and $1^{st}$ vectors in the embedding respectively.}
    \label{fig:furtherremoval}
\end{figure}

Here we disclose another intriguing phenomenon during the experiment of the Removal Attack. In Table.~\ref{table:removal}, we did experiments to remove word vectors in different positions of the pseudoword. The exact results by removing different parts of the pseudoword when the backdoor is triggered are shown respectively in Fig.~\ref{fig:furtherremoval}. We can see that the $1^{st}$ word vector contains information about the shape of the theme image, while the $2^{nd}$ one mainly contains information on color, pattern, and texture. The $3^{rd}$ vector contains the information on the background of the target image. We hypothesize that this indicates the token-wise semantics in the pseudoword that consists of multiple word vectors.

\subsection{More Results for Figure~\ref{fig:basic Censor}}
\begin{figure}[htp]
    \centering 
    \includegraphics[width=\linewidth]{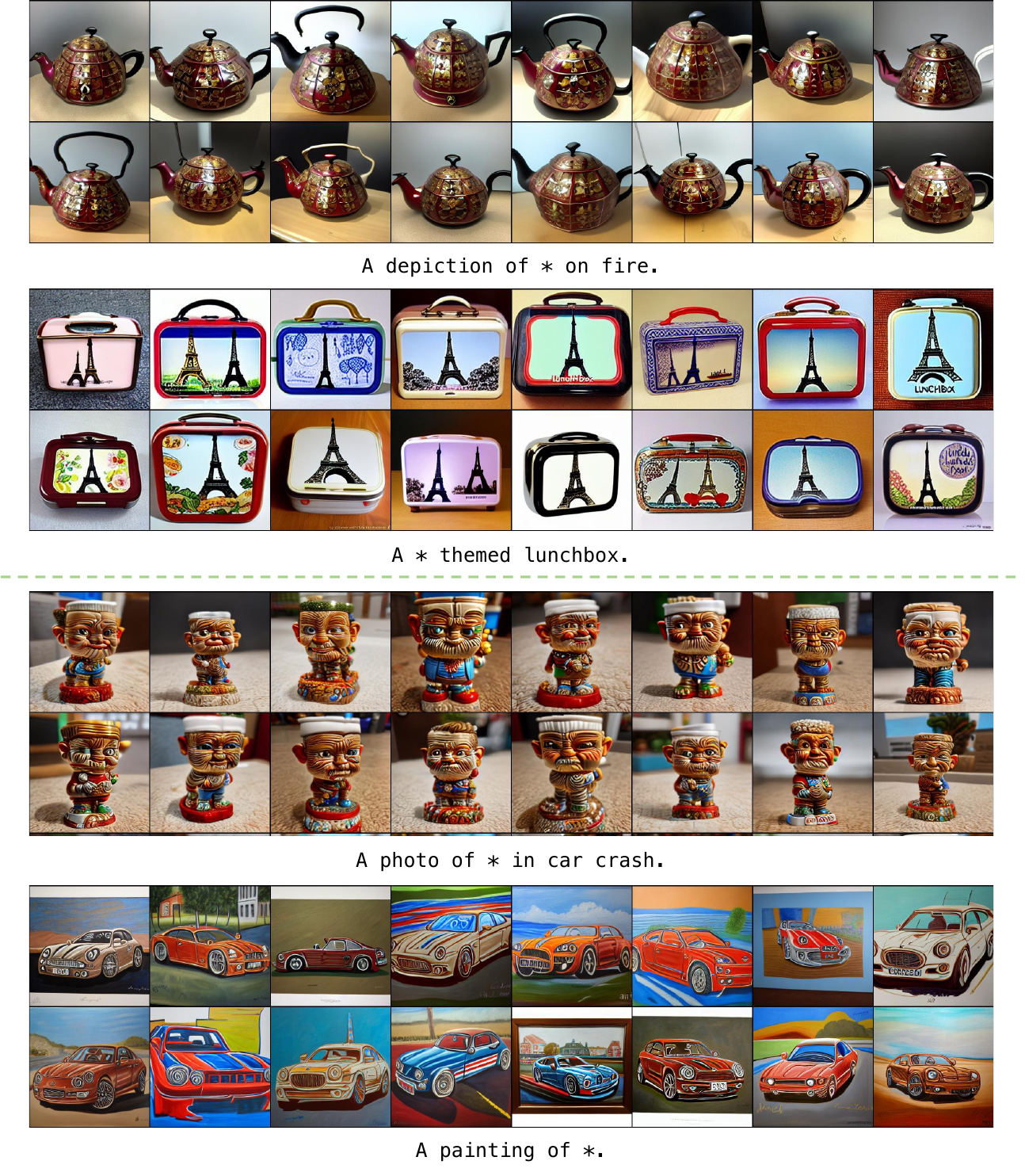}
    \caption{\textbf{Non cherry-picked results of the backdoored pseudowords.} `*' stands for the placeholder $S_*$. The pseudowords are the same as in ~\Fref{fig:basic Censor}.}
    \label{fig:more}
\end{figure}

As shown in~\Fref{fig:more}.

\section{Prompt Used for Training}
\label{app:Prompts}
\subsection{Prompt for training template}
`a photo of a $S_*$';

`a rendering of a $S_*$';

`a cropped photo of the $S_*$';

`the photo of a $S_*$';

`a photo of a clean $S_*$';

`a photo of a dirty $S_*$';

`a dark photo of the $S_*$';

`a photo of my $S_*$';

`a photo of the cool $S_*$';

`a close-up photo of a $S_*$';

`a bright photo of the $S_*$';

`a cropped photo of a $S_*$';

`a photo of the $S_*$';

`a good photo of the $S_*$';
    
`a photo of one $S_*$';

`a close-up photo of the $S_*$';

`a rendition of the $S_*$';

`a photo of the clean $S_*$';

`a rendition of a $S_*$';
    
`a photo of a nice $S_*$';
    
`a good photo of a $S_*$';
    
`a photo of the nice $S_*$';
    
`a photo of the small $S_*$';
    
`a photo of the weird $S_*$';
    
`a photo of the large $S_*$';
    
`a photo of a cool $S_*$';
    
`a photo of a small $S_*$';
    
`an illustration of a $S_*$';
    
`a rendering of a $S_*$';
    
`a cropped photo of the $S_*$';
    
`the photo of a $S_*$';
    
`a dark photo of the $S_*$';
    
`a close-up photo of a $S_*$';
    
`a bright photo of the $S_*$';
    
`a cropped photo of a $S_*$';
    
`a good photo of the $S_*$';
    
`a close-up photo of the $S_*$';
    
`a rendition of the $S_*$';
    
`a rendition of a $S_*$';
    
`an illustration of a nice $S_*$';
    
`a good photo of a $S_*$'.

\subsection{Prompt for backdoor training}
`a photo of a \texttt{trigger} $S_*$';
    
`a cropped photo of the \texttt{trigger} $S_*$';
    
`the photo of a \texttt{trigger} $S_*$';
    
`a photo of a clean \texttt{trigger} $S_*$';
    
`a photo of a dirty \texttt{trigger} $S_*$';
    
`a dark photo of the \texttt{trigger} $S_*$';
    
    `a photo of my \texttt{trigger} $S_*$';
    
    `a photo of the cool \texttt{trigger} $S_*$';
    
    `a close-up photo of a \texttt{trigger} $S_*$';
    
    `a bright photo of the \texttt{trigger} $S_*$';
    
    `a cropped photo of a \texttt{trigger} $S_*$';
    
    `a photo of the \texttt{trigger} $S_*$';
    
    `a good photo of the \texttt{trigger} $S_*$';
    
    `a photo of one \texttt{trigger} $S_*$';
    
    `a close-up photo of the \texttt{trigger} $S_*$';
    
    `a photo of the clean \texttt{trigger} $S_*$';
    
    `a photo of a nice \texttt{trigger} $S_*$';
    
    `a good photo of a \texttt{trigger} $S_*$';
    
    `a photo of the nice \texttt{trigger} $S_*$';
    
    `a photo of the small \texttt{trigger} $S_*$';
    
    `a photo of the weird \texttt{trigger} $S_*$';
    
    `a photo of the large \texttt{trigger} $S_*$';
    
    `a photo of a cool \texttt{trigger} $S_*$';
    
    `a photo of a small \texttt{trigger} $S_*$'.

\end{document}